\documentclass[%
 reprint,
 longbibliography,
 amsmath,amssymb,
 aps,
]{revtex4-2}

\usepackage[pdftex,colorlinks=true,linkcolor=blue,bookmarks=false,citecolor=blue,urlcolor=blue,hypertexnames=false]{hyperref}

\usepackage{graphicx}
\usepackage{dcolumn}
\usepackage{bm}
\usepackage{hyperref}
\usepackage[mathlines]{lineno}
\usepackage{braket}
\usepackage{svg}
\usepackage{xcolor}
\definecolor{crimson}{rgb}{0.7, 0.1, 0.1}

\newcommand{\rr}[1]{\textcolor{black}{#1}} 
\newcommand{\yb}[1]{\textcolor{black}{#1}} 
\newcommand{\ad}[1]{\textcolor{black}{#1}} 
\newcommand{\dw}[1]{\textcolor{black}{#1}} 

\usepackage{comment}
\usepackage{multirow}
\usepackage{graphicx}
\usepackage{dcolumn}
\begin{document}

\title{Exploring light-induced phases of 2D materials in a modulated 1D quasicrystal} 

\author{Yifei Bai}
\author{Anna R. Dardia}
\author{Toshihiko Shimasaki}
\author{David M. Weld}
\email{weld@ucsb.edu}
\affiliation{%
Department of Physics, University of California, Santa Barbara, California 93106, USA
}%

\date{\today}

\begin{abstract}
Light-induced quantum phases offer the potential for simple and powerful tuning of material properties.
For example, simply illuminating \ad{2D materials in the integer quantum Hall regime with polarized light is predicted to drive quantum phase transitions. }
\dw{Such phenomena are largely beyond the current frontier of solid state experiments due to technical limitations on laser intensity and material purity}. However, the Harper-Hofstadter mapping which relates a two-dimensional integer quantum Hall system to a 1D quasicrystal enables the same polarization-dependent light-induced phase transitions to be observed using a quantum gas in a driven quasiperiodic optical lattice. \rr{We report results of such an experiment.} 
We observe an interlaced phase diagram of localization-delocalization phase transitions as a function of drive polarization and amplitude. Elliptically polarized driving can stabilize an extended critical phase featuring multifractal wavefunctions; we observe signatures of this phenomenon in \dw{anomalous polarization-dependent} subdiffusive transport. In this regime, increasing the strength of the quasiperiodic potential can enhance rather than suppress transport. These experiments demonstrate a simple method for synthesizing exotic multifractal states and exploring light-induced quantum phases across different dimensionalities. 

\end{abstract}

\maketitle

\begin{figure*}[ht!]
    \centering
    \includegraphics[width=1\linewidth]{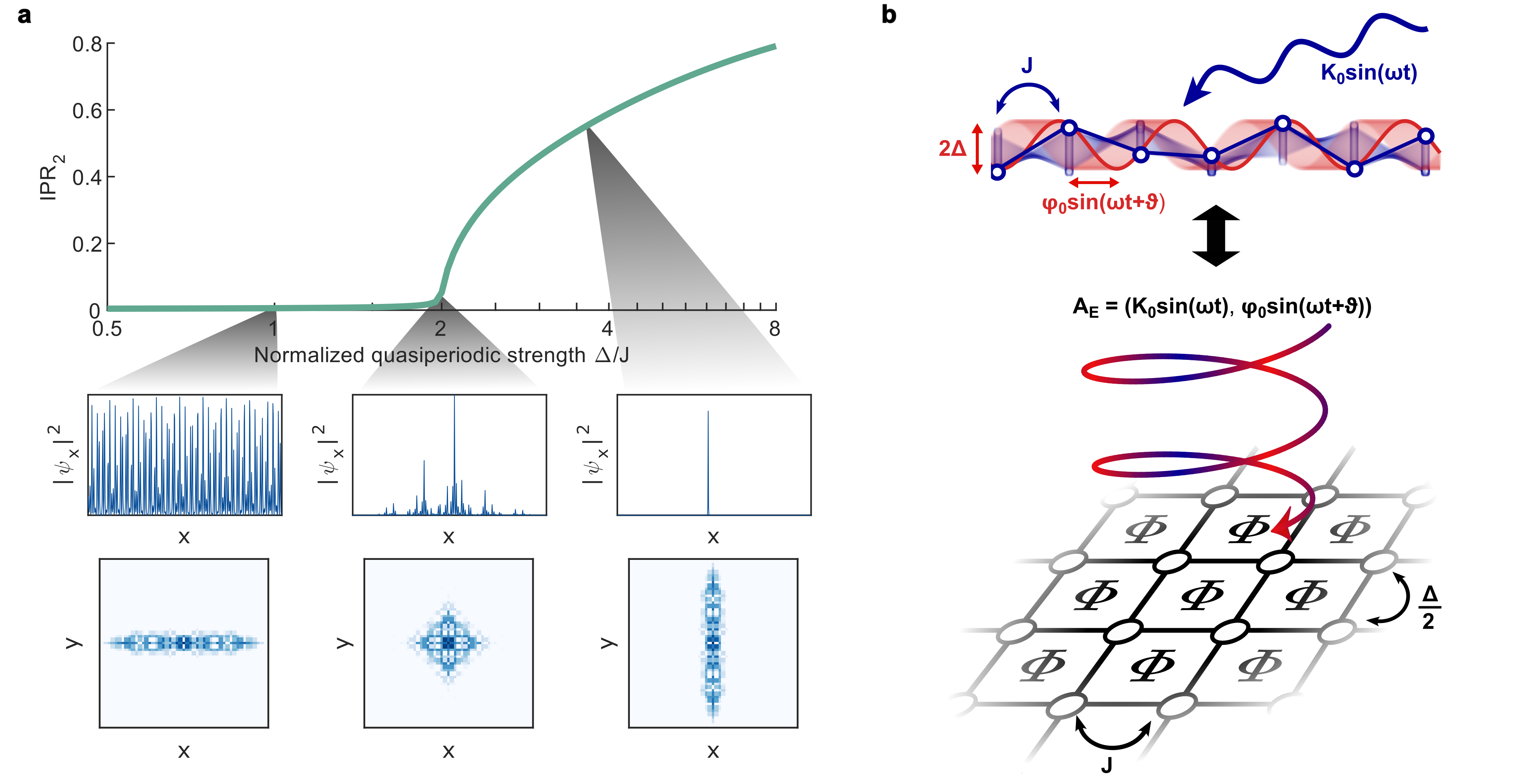}
    \caption{
    \rr{
    \textbf{Correspondence between modulated 1D quasicrystal and optically-driven 2D crystal}. \textbf{a,} The undriven 1D AAH system undergoes a localization phase transition as the quasiperiodic strength $\Delta$ increases, diagnosed by the inverse participation ratio (IPR) and the density profiles of a representative eigenstate. The localization phase transition in 1D is mapped to a 90${}^\circ$ rotation of directional localization in 2D, as shown by the calculated time-evolved 2D density profile. \textbf{b,} The 1D quasicrystal is subjected to both dipolar and phasonic modulation. In 2D, the quasiperiodic strength $\Delta$ becomes the tunneling along the additional dimension and thus controls the tunneling anisotropy 
    see also Appendix \ref{apd:HHmap}). 
    The dipolar and phasonic modulations map to orthogonal components of the vector potential of the irradiation in 2D. 
    }   
    }
    \label{fig:setup}
\end{figure*}

\begin{figure*}[th!]
    \centering
    \includegraphics[width=\linewidth]{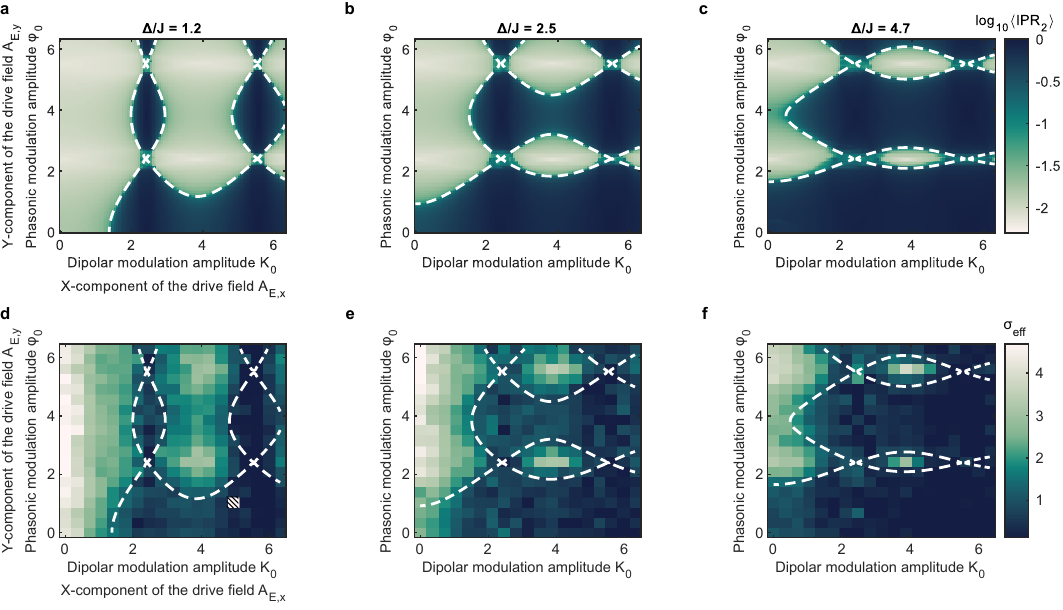}
    \caption{\textbf{Interlaced localization-delocalization quantum phase transitions controlled by polarized driving}. Figure shows theoretically predicted and experimentally measured phase diagrams at quasiperiodic strengths $\Delta/J \approx 1.2$, $2.5$ and $4.7$; below, near and above the critical strength without driving. Axes represent dipolar and phasonic components of the drive in 1D, corresponding to $x$ and $y$ components of the driving electric field in 2D. \textbf{a-c} Theoretically calculated inverse participation ratio (IPR) averaged over all Floquet eigenstates in the central Brillouin zone. System size is 200 lattice sites. \textbf{d-f} Measured effective fractional width $\sigma_\mathrm{eff}$ after $t_\mathrm{hold} = 1.25 \, \mathrm{s} \approx 400 \, T_J$. The patched data point in \textbf{(d)} is cut due to significant atom loss. All data, theoretical or experimental, are taken with 
    \yb{$\vartheta=\pi/2$} (elliptical polarization).}
    \label{fig:PDs}
\end{figure*}

Tuning quantum matter by adjusting the strength, frequency, and polarization of 
external irradiation is a powerful and flexible method for creating novel on-demand properties, as demonstrated by rapid developments in the field of strongly driven solids~\cite{RevModPhys.93.041002,Weitenberg_Simonet_2021_FlqReview, Basov_Averitt_Hsieh_OnDemand_2017, Wang_Steinberg_Jarillo-Herrero_Gedik_2013, Uzan-Narovlansky_interbandBerryPhase, RevModPhys.93.041002}. 
In such experiments the polarization of the driving waveform is a crucial degree of freedom; for example, it can control time-reversal symmetry, stabilizing phases such as Floquet topological insulators \cite{Kitagawa_PhysRevB.84.235108, Wang_Steinberg_Jarillo-Herrero_Gedik_2013}. 
Realizing the full promise of driven condensed matter is a serious challenge; in addition to their intrinsic complexity, these experiments require optical intensities and pulse times at the edge of feasibility with current materials and laser technology~\cite{Basov_Averitt_Hsieh_OnDemand_2017}. Complementary input from experiments on cold atom platforms, at very different energy scales but realizing the same Hamiltonians \rr{\cite{Lignier_DL_PhysRevLett.99.220403, Arlinghaus_Holthaus_2010_OLstrongfieldSim}}, can play a key role in exploring the frontiers of driven matter and guiding future experiments on quantum materials under intense arbitrarily-polarized irradiation.

In this manuscript we describe and demonstrate an experiment along these lines, using cold atoms in a doubly-driven 1D quasicrystal to explore the phase diagram of 
\rr{2D Bloch electrons in a high magnetic field} 
 illuminated with 
intense light of tunable polarization \cite{Bai_PhysRevB.111.115163}. This depends on a cross-dimensional correspondence mapping a 1D quantum quasicrystal to and from a 2D electron gas in the integer quantum Hall regime~\cite{harper_single_1955, Hofstadter_OG_PhysRevB.14.2239, kraus_quasiperiodicityTranscend_2016, Jagannathan}, in which the quasicrystal encodes the additional dimension in its phasonic degree of freedom~\cite{kraus_quasiperiodicityTranscend_2016, Aubry_1980}.
The Aubry-André localization phase transition in 1D \cite{aubry1980analyticity} corresponds to a 90-degree rotation of directional conductivity in the 2D model (Fig.~\ref{fig:setup}\rr{(a)}) \cite{Hofstadter_OG_PhysRevB.14.2239, Barelli_PhysRevLett.83.5082, Sun_Ralston_1991,Itzler_Bojko_Chaikin_1992, Itzler_Danner_Bojko_Chaikin_1994, supp_mat}, which was recently proposed to be potentially observable in strained Moiré materials \rr{even in the presence of interactions}~\cite{Paul_PhysRevLett.132.246402, Paul_Crowley_2024_interaction}.
This mapping, which is extended here to a time-dependent context, also enables the counter-intuitive realization of polarized driving in one dimension. 
A time-dependent dipolar electric field in the 1D quasicrystal maps to an electric field along one dimension in the 2D Hofstadter model, and a time-dependent phasonic mode in the quasicrystal maps to an electric field along the other dimension in 2D (Fig.~\ref{fig:setup}\rr{(b)}). The combination of both modulations with variable amplitude and relative phase in 1D then directly corresponds to tunably polarized laser illumination of the 2D \ad{Hofstadter model}. The  \dw{experiment} is capable of exploring polarization-dependent light-induced quantum phase transitions at extremely strong effective drive amplitudes.

We experimentally measure the effects of strong polarized driving  by observing the expansion of a quantum gas in a doubly-driven 1D quasiperiodic optical lattice. The results reveal an intricate tessellated metal-insulator quantum phase diagram controlled by the relative strengths and phases of the two driving fields, which can be understood from the 2D perspective as a result of competition between drive-induced dynamic localization \cite{Lignier_DL_PhysRevLett.99.220403, Shimasaki_PhysRevLett.133.083405} along different dimensions \cite{Shimasaki_PhysRevLett.133.083405, Bai_PhysRevB.111.115163}, leading to coherent switching of directional localization by light. The polarization of the driving field in the higher-dimensional space plays a crucial and nontrivial role: for example, we observe that an elliptically polarized field 
\rr{produces anomalous transport attributable to eigenstate multifractality, whereas a linearly polarized field of the same intensity does not. This anomalous transport is counter-intuitively enhanced rather than suppressed by the quasiperiodic potential}. While  \rr{non-interacting} multifractal quantum states have recently attracted intense theoretical interest ~\cite{Liu_LSP_10.21468/SciPostPhys.12.1.027,Wang_Raman-Critical_PhysRevLett.125.073204, Gonifmmode_CP_theory_PhysRevLett.131.186303, Gonifmmode_QP_ergodicity_PhysRevB.108.104201, zhou2025fundamentallocalizationphasesquasiperiodic, Bai_PhysRevB.111.115163,shimasaki2022anomalous, XIAO_Critical_MomentumLattice_20212175, Li_modifiedAAH_exp_2023, huang2025exactquantumcriticalstates, Gonçalves_2024NatPhys}, 
\yb{experimental observations have been limited to date by the need for fine-tuned Hamiltonians~\cite{Wang_Raman-Critical_PhysRevLett.125.073204, huang2025exactquantumcriticalstates, XIAO_Critical_MomentumLattice_20212175}. The prospect of Floquet-engineering multifractal phases has been discussed and debated for decades~\cite{Ketzmerick_Kruse_Geisel_1999_Lanzcos, Borgonovi_Shepelyansky_1995_KHM, Prosen_dimer_PhysRevLett.87.066601, Roy_MFwithoutFineTune_10.21468/SciPostPhys.4.5.025, Gonifmmode_QP_ergodicity_PhysRevB.108.104201, Bai_PhysRevB.111.115163, shimasaki2022anomalous}. Our results constitute experimental evidence that multifractality can arise simply and without any fine tuning, under a range of amplitudes of elliptically polarized illumination.
}

Each experiment begins with a ${}^{84}\mathrm{Sr}$ Bose-Einstein condensate adiabatically loaded into a combined crossed optical dipole trap and 1D quasiperiodic bichromatic lattice. The dipole trap is suddenly switched off and sinusoidal modulation of the two optical lattices comprising the quasiperiodic lattice is simultaneously switched on, allowing the condensate to evolve in the modulated quasicrystal potential.   
The deeper primary lattice of depth $V_P$  provides lateral confinement against gravity and admits a tight-binding description, while the secondary lattice with depth $V_S \ll V_P$ creates spatial quasiperiodicity \rr{\cite{Modugno_2009}}. Each lattice is formed by two counterpropagating beams which can be individually frequency-modulated. Frequency modulation of one beam in the primary lattice creates an inertial force on the atoms which mimics the dipolar force on charges in an oscillating electric field. Relative motion between the two lattices introduces a phasonic modulation \cite{Shimasaki_PhysRevLett.133.083405, Bai_PhysRevB.111.115163, supp_mat}. 
The resulting Hamiltonian in the velocity gauge is \cite{supp_mat} 
\begin{align}
    \hat H (t) = &\sum_j -J e^{-iK_0 \sin (\omega t)}\hat c_{j+1}^\dagger \hat c_j + \mathrm{h.c.} \nonumber \\
    & + \Delta \cos(2\pi\beta j + \kappa + \varphi_0\sin(\omega t + \vartheta) ) \hat n_j\,, \label{eq:DDAAH_comoving}
\end{align}
where $J$ is the tunneling strength, $\Delta$ is the quasiperiodic strength, and $\beta \approx 1.2165$ is the incommensurate ratio set by the ratio of lattice constant of the two optical lattices. \rr{This Hamiltonian neglects interparticle interactions, and we have verified experimentally that atom-atom interactions play a negligible role in the phenomena we report~\cite{supp_mat}.}
We set $V_P = 9 E_\mathrm{R,P}$ with $E_\mathrm{R,P} \approx h \times 2.1\,\mathrm{kHz}$ the recoil energy from the primary lattice, leading to a tunneling time $T_J = \hbar/J = 3.1\,\mathrm{ms}$ and a tunneling strength $J \approx h\times 50.9\,\mathrm{Hz}$. 
$K_0$ is the dipolar modulation amplitude and $\varphi_0$ is the phasonic modulation amplitude; $\omega = 2\pi\times 500 \, \mathrm{Hz} \approx 9.8J/\hbar$ is the modulation frequency, which is within the first gap of the unperturbed primary lattice and above the total bandwidth of the model \cite{Sun_optimalFreqWindow_PhysRevResearch.2.013241}. At zero drive amplitude $K_0=\varphi_0=0$ the Hamiltonian reduces to the well-known Aubry-André model~\cite{aubry1980analyticity},
\yb{also corresponding to one Fourier slice of the 2D Harper-Hofstatder model \cite{harper_single_1955, Hofstadter_OG_PhysRevB.14.2239, Barelli_PhysRevLett.83.5082} with the 2D annihilation (creation) operators defined through the Fourier transform $\hat c_{j, \kappa}^{(\dagger)} = \sum_l e^{-i \kappa l}\hat d_{j,l}^{(\dagger)}$}. 
It features a localization quantum phase transition at $\vert \Delta/J\vert =2$. Finite-strength dipolar and phasonic driving effectively rescale $J$ and $\Delta$ respectively, and can thereby move the quantum phase transition in either direction \cite{Shimasaki_PhysRevLett.133.083405, Dotti_doubleLoc_PhysRevResearch.7.L022026, Bai_PhysRevB.111.115163}.

A crucial and unique feature of this experiment is the ability to control the temporal phase difference $\vartheta$ between the dipolar and phasonic modulations. From the perspective of the cross-dimensional mapping, this temporal phase difference maps exactly to the orientation angle of the polarization ellipse in 2D \cite{Bai_PhysRevB.111.115163, supp_mat}. Thus, we can fully control the state of the polarization of the irradiation: applying linear, circular, or elliptical polarization in the 2D space is simply a matter of selecting the appropriate amplitudes $K_0$ and $\varphi_0$ and the phase $\vartheta$. 

Our main observable is the expansion of the atoms as measured by the time-dependent width $\sigma_x (t)$, extracted by fitting the density distribution to a \rr{bimodal Gaussian (see Appendix \ref{apd:methods}}). We define the fractional expansion $\sigma:= (\sigma_x-\sigma_0)/\sigma_0$ where $\sigma_0 \approx 10\,\mu \mathrm{m}$ is the initial width of the condensate, to account for the small but measurable dependence of the initial width on the quasiperiodic strength $\Delta$ \cite{supp_mat}. In a localized phase, $\sigma \approx 0$, while $\sigma$ increases with time in a delocalized phase. To characterize the phase diagram of the Hamiltonian with experiments with fixed evolution time, we further define an effective fractional expansion $\sigma_\mathrm{eff} = \sigma/\max{(\vert\mathcal{J}_0(K_0)\vert, \langle\delta\sigma\rangle})$ which normalizes with respect to the Bessel-modified hopping rate \cite{Lignier_DL_PhysRevLett.99.220403}. Here $\langle\delta\sigma\rangle \approx 0.12$ is the standard deviation of repeated measurements, averaged over the entire phase diagram.

The first main result of this work is the experimental measurement of the phase diagram of Hamiltonian (\ref{eq:DDAAH_comoving}) as a function of both the dipolar and phasonic modulation amplitude. Fig. \ref{fig:PDs} shows predicted and measured phase diagrams at different quasiperiodic potential strengths $\Delta$ after the same evolution time $t_\mathrm{hold} = 1.25 \, \mathrm{s} \approx 400 \, T_J$. Note that the color axes represent \rr{distinct but complementary} quantities for the theoretical and experimental diagrams: inverse participation ratio (IPR) of the Floquet eigenstates in the position basis $\sum_j \vert \psi_j \vert^4$ \cite{supp_mat} is the numerically accessible \rr{spectral} diagnostic for localization, while $\sigma_\mathrm{eff}$ is the experimentally measurable \rr{dynamical} one. 
\rr{The IPR and the real-space expansion provide complementary probes of the localization phase diagram, and Fig.~2 demonstrates the agreement between them.} The competition between the dipolar and phasonic modulations in 1D, and thus between the $x$ and $y$ components of the drive field in 2D, leads to tessellated phase diagrams featuring interlaced regions of localized and delocalized phases. From the 2D perspective, these results demonstrate that intense polarized irradiation rescales the tunneling anisotropy, coherently switching the axis of conductivity. 

The main features of the measured phase diagram can be understood analytically. Averaged over one drive cycle, the dipolar drive rescales the effective tunneling strength by a zeroth-order Bessel function: $J_\mathrm{eff}=\langle J e^{-iK_0 \sin (\omega t)}\rangle_T = J \mathcal{J}_0 (K_0)$. At the Bessel zeros where $J_\mathrm{eff} = 0$, the resulting dynamic localization \cite{Lignier_DL_PhysRevLett.99.220403} freezes transport along the physical dimension and thus favors a localized phase  \cite{Dotti_doubleLoc_PhysRevResearch.7.L022026}. Conversely, the phasonic drive rescales the effective quasiperiodic strength according to $\Delta_\mathrm{eff} = \Delta \mathcal{J}_0 (\varphi_0)$, and thus favors a delocalized phase 
\cite{Shimasaki_PhysRevLett.133.083405}. The cross-dimensional mapping makes an exact analogy to dynamic localization apparent: in the 2D space, the quasiperiodic strength $\Delta$ is the tunneling and $\varphi (t)$ is the field component along the additional dimension (Fig. \ref{fig:setup}\rr{(b)}). The position of the quantum phase transition in the $K_0-\varphi_0$ plane thus satisfies $\Delta \vert\mathcal{J}_0(\varphi_0)/\mathcal{J}_0(K_0) \vert = 2J$; these are the boundary lines drawn on Fig. \ref{fig:PDs}.

\begin{figure}[t!]
    \centering
    \includegraphics[width=\linewidth]{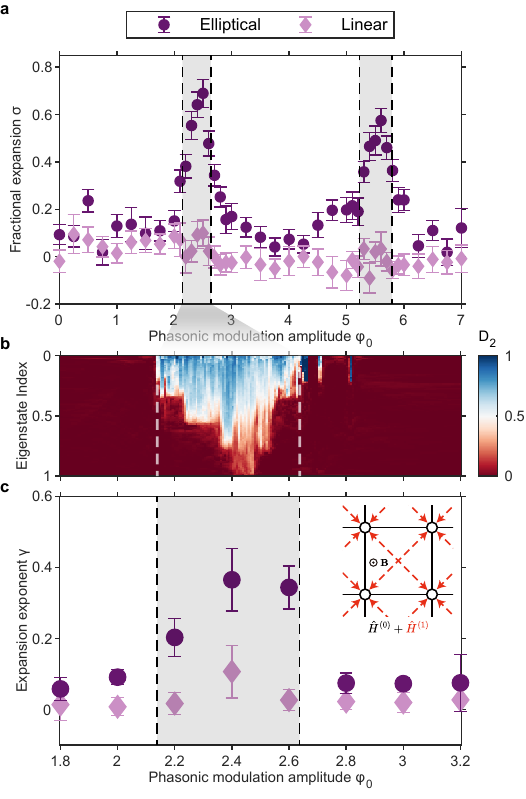}
    \caption{\textbf{Elliptically polarized driving induces exotic transport.} Phasonic driving modifies the dynamically localized state ($K_0\approx 2.405$) in a polarization-dependent way. \textbf{a,} Fractional expansion $\sigma$ as a function of phasonic modulation amplitude $\varphi_0$, for elliptical (circles) or linear (diamonds) polarization. For these data $t_\mathrm{hold} = 10 \, \mathrm{s}$ and $\hbar\omega = h\times 500\, \mathrm{Hz} \approx 9.8J$. 
     \yb{Dashed lines and shaded areas represent the theoretically predicted phase transition into a multifractal phase $\vert\Delta_\mathrm{eff}\vert = 2\vert J_\mathrm{EP} \vert$ (see Eq.\ref{eq:H1} and \cite{Han_HarperNNN_PhysRevB.50.11365}, and discussion in the text) for elliptically polarized driving.}
    Error bars represent standard error of the mean (s.e.m.) of repeated measurements. \textbf{b,} Theoretically calculated fractal dimension $D_2$ of all eigenstates, zoomed in to the region $\varphi_0\in [1.8,3.2]$. \textbf{c,} Measured expansion exponents for elliptically and linearly polarized driving, using the same experimental parameters. Error bars represent 95\% confidence interval of the fit. Inset is a visualization of the effective Hamiltonian under elliptical polarization in 2D, where red dashed arrows represent NNN hopping (Eq. \ref{eq:H1}). }  
    \label{fig:pol_induced_dyn}
\end{figure}

Tuning the polarization of the driving radiation reveals additional rich physics~\cite{Bai_PhysRevB.111.115163}. In the second main result of this work, we experimentally measure and theoretically characterize the nontrivial effects of varying the drive polarization. The polarization is straightforwardly tuned by adjusting the phase difference $\vartheta$ between dipolar and phasonic modulations. In practice, this is implemented by delaying the modulation of the secondary lattice by an amount of time $\tau_\mathrm{delay}$. For elliptically polarized irradiation in 2D 
\yb{($\vartheta =\pm\pi/2$)}, $\tau_\mathrm{delay} = 0$, while for linearly polarized irradiation 
\yb{($\vartheta = 0$)}, $\tau_\mathrm{delay} > 0$ is completely determined by the desired dipolar and phasonic modulation amplitudes $(K_0, \varphi_0)$ and the driving frequency $\omega$~\cite{supp_mat}. 

As a starting point for these investigations, we consider the dynamically localized strip where $K_0=K_0^{\mathrm{DL}} = 2.405$. Here $J_\mathrm{eff} \approx 0$: kinetic energy is effectively eliminated to zeroth order, so naive theory predicts a localized phase for essentially all values of the phasonic drive amplitude $\varphi_0$ since $\Delta_\mathrm{eff}/J_\mathrm{eff}\rightarrow \infty$. In accordance with this expectation, for linearly polarized modulation we measure barely any expansion at any $\varphi_0$ even after $t_\mathrm{hold} = 10\,\mathrm{s} \approx 3200 T_J$ (diamonds in Fig.~\ref{fig:pol_induced_dyn}(a)). However, when we apply elliptically polarized modulation with exactly the same dipolar and phasonic amplitudes, the results differ drastically, in a way which disagrees sharply with the naive expectation of featureless localization. In particular, we observe peaks in the expansion rate when $\varphi_0$ is near the zeros of $\mathcal{J}_0 (\varphi_0)$, at $\varphi_0 \approx 2.405$ and $5.520$ \rr{where the time-averaged Hamiltonian vanishes completely.}

What drives this clear difference in transport between the two drive polarizations? To probe this question experimentally, we focus on the first delocalized region around $\varphi_0 \approx 2.405$ and measure the expansion exponent $\gamma$ that 
\yb{quantifies and connects the transport dynamics to the fractal dimension}
\cite{Ketzmerick_spreading_PhysRevLett.79.1959}. This is extracted by fitting the \rr{width} $\sigma_x(t)$ to the phenomenological relation $\sigma_x(t)/\sigma_0 = (1+t/t_0)^\gamma$ where $t_0$ represents an ``activation time" \cite{Lucioni2011_subdiffMF}. We use $t_\mathrm{hold}$ values up to $20\,\mathrm{s}$ ($\approx 6400 T_J$) to minimize the effects of initial state dependence. In a localized phase, $\gamma \approx 0$; in a ballistic phase, $\gamma\approx 1$; for diffusive dynamics, $\gamma = 0.5$. As shown in \ref{fig:pol_induced_dyn}(c), the measured exponents $\gamma$ are always close to zero for the linearly polarized driving case, corresponding to a localized phase. In the case of elliptically polarized driving, however, the exponents are nonvanishing within a finite window of phasonic amplitudes $\varphi_0$ near 2.4. For $\varphi_0$ outside this region, the exponent drops close to zero. 
The fact that both measurements are taken with identical drive amplitudes $K_0$ and $\varphi_0$ suggests that the observed subdiffusive expansion is not due to mean-field interaction \cite{Lucioni2011_subdiffMF} or transverse heating \cite{Chaudhury_transverseInstability_PhysRevA.91.023624, Reitter_heating_PhysRevLett.119.200402}, but rather originates from the spectral structure \cite{Ketzmerick_spreading_PhysRevLett.79.1959}. We interpret these results as evidence for the presence of a subdiffusive delocalized phase which emerges only in the presence of elliptically polarized driving. 

This interpretation is also supported by further theoretical analysis. Motivated by the connection between the expansion exponent and eigenstate fractal dimension \cite{Ketzmerick_spreading_PhysRevLett.79.1959}, we compute the fractal dimension $D_2$ of the \yb{Floquet} eigenstates of Hamiltonian (\ref{eq:DDAAH_comoving}) under the condition of elliptically polarized modulation. The results are shown in Fig. \ref{fig:pol_induced_dyn}(b). For ballistic (localized) phases, $D_2 = 1 \; (0)$ for all eigenstates and there is no fractality. In contrast, we calculate that $D_2 \neq 0,1$ for some eigenstates in a window of $\varphi_0$ corresponding exactly to the window where we observe nontrivial expansion exponents. These nontrivial values of $D_2$ indicate eigenstate multifractality without fine-tuning to a critical point.
\rr{Our calculation suggests the presence of a multifractal phase: a nonvanishing fraction of eigenstates are multifractal over a range of parameters rather than just at one critical point ~\cite{Han_HarperNNN_PhysRevB.50.11365, Chang_MFanalysis_modifiedAAH_PhysRevB.55.12971, Avila2017_proof_CP}.}

\begin{figure}[t]
    \centering
    \includegraphics[width=\linewidth]{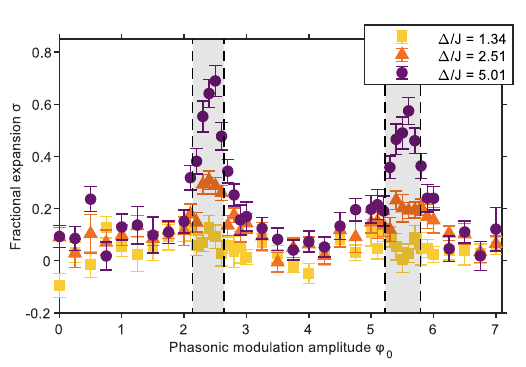}
    \caption{\textbf{Anomalous expansion under elliptically polarized modulation is enhanced by the quasiperiodic strength}. As the quasiperiodic strength $\Delta$ is increased, the expansion increases in the shaded region representing a multifractal phase, but remains localized otherwise. Here the condensate expands for $t_\mathrm{hold} = 10\, \mathrm{s}$ for all data sets. Error bars represent s.e.m.~of repeated measurements. }
    \label{fig:AnomExpansion}
\end{figure}

An especially clear picture of this observation emerges in the 2D perspective. As has been known for more than a decade from the study of Floquet topological insulators \cite{Kitagawa_PhysRevB.84.235108}, elliptically polarized irradiation in 2D can drive effective next-nearest-neighbor (NNN) hopping in real space due to time-reversal symmetry breaking, whereas linearly polarized irradiation cannot. For our rectangular Hofstadter model, only elliptically polarized radiation can generate an effective NNN hopping across the diagonal of a plaquette. It has long been known that such a beyond NN hopping can lead to a broad multifractal phase \cite{Han_HarperNNN_PhysRevB.50.11365, Chang_MFanalysis_modifiedAAH_PhysRevB.55.12971, Avila2017_proof_CP}.
This theoretical picture supports the conclusion that we can engineer multifractal phases by applying elliptically polarized irradiation in the presence of irrational flux per plaquette \cite{Bai_PhysRevB.111.115163}. Quantitatively, these predictions can be derived from a high-frequency expansion \cite{Goldman_PhysRevX.4.031027, Eckardt_2015_HFE} that leads to the effective Hamiltonian $\hat H_\mathrm{eff} \approx \hat H^{(0)} + \hat H^{(1)}$, whose 2D correspondence is schematically shown in the inset of Fig.\ref{fig:pol_induced_dyn}(c). For $K_0 = K_0^\mathrm{DL}$, $\hat H^{(0)}$, the time-averaged term, is just the quasiperiodic potential dressed by the phasonic modulation, since $J_\mathrm{eff} = 0$. For linearly polarized driving, $\hat H^{(1)} = 0$. For elliptically polarized driving corresponding to Fig. \ref{fig:pol_induced_dyn},
\begin{equation}
    \hat H^{(1)} = J_\mathrm{EP}(\varphi_0)\sum_j \cos \left(2\pi\beta j + \pi\beta + \kappa \right)\hat c_{j+1}^\dagger \hat c_j+ \mathrm{h.c.} \label{eq:H1} 
\end{equation}
where the tunneling energy scale is 
\begin{multline}
        J_\mathrm{EP}(\varphi_0) = \frac{4J \Delta}{\hbar \omega}\sin(\pi\beta) \\
        \times \sum_{n=0}^\infty \frac{(-1)^{n+1}}{2n+1} \mathcal{J}_{2n+1} (K_0^{\mathrm{DL}})\mathcal{J}_{2n+1} (\varphi_0), \label{eq:j1}    
\end{multline}
The full expression of $\hat H^{(1)}$ in a general setting can be found in \cite{Bai_PhysRevB.111.115163} and the supplementary information \cite{supp_mat}. The elliptically polarized drive thus leads to 
spatially quasiperiodic tunneling which is projected from the NNN tunneling in 2D \cite{Han_HarperNNN_PhysRevB.50.11365} \rr{and which creates a multifractal delocalized phase}. This expression allows us to quantitatively predict the phase boundary between a localized and a critical phase at $\vert\Delta_\mathrm{eff}\vert = 2\vert J_\mathrm{EP}(\varphi_0)\vert$ \cite{Han_HarperNNN_PhysRevB.50.11365}. The resulting phase boundaries shown in Figs. 3 and \ref{fig:AnomExpansion} demonstrate excellent agreement with the experimental measurements of width $\sigma$ and exponent $\gamma$, and with the predicted multifractal regime of non-integer $D_2$. Taken together, these results thus provide strong evidence for multifractality-driven anomalous transport due solely to the polarization of the applied drive. 
\yb{We note that the experimental conditions and physical mechanism of the observed phasonically driven delocalization are completely different from those observed previously in \cite{Shimasaki_PhysRevLett.133.083405}. The similarity in the functional shape is simply because the delocalized phases, ballistic or multifractal, are around the zeros of the Bessel function $\mathcal{J}_0(\varphi_0)$. The delocalization observed in \cite{Shimasaki_PhysRevLett.133.083405}, corresponding to $K_0 = 0$ and thus linearly polarized modulation, is driven by the rescaling of $\Delta_\mathbf{eff}$ and does not involve any underlying eigenstate multifractality.}

A striking further prediction of the theory described in Eqs.~(\ref{eq:H1}) and (\ref{eq:j1}) is that the anomalous expansion driven by elliptically polarized irradiation should be enhanced, rather than suppressed, for larger values of the quasiperiodic potential strength $\Delta$. This counter-intuitive prediction, while not obvious from the time-dependent Hamiltonian itself, naturally emerges from the effective Hamiltonian description, since $\hat H^{(1)}$ resembles a quasiperiodically-modulated kinetic energy term whose strength is proportional to $\Delta$. 
The third and final main result of this work is the experimental confirmation of this prediction, shown in Fig.~\ref{fig:AnomExpansion}. The two transport peaks corresponding to the anomalous phase become much more pronounced as quasiperiodic strength increases, from being barely visible at $\Delta/J \approx 1.34$ to very prominent at $\Delta/J \approx 5.01$. This demonstrates the predicted enhancement of anomalous transport by the quasiperiodic potential. From the 2D perspective, increasing $\Delta$, while suppressing the transport along the physical dimension, enhances the strength of NNN hopping that supports the anomalous multifractal dynamics (see inset of Fig.\ref{fig:pol_induced_dyn}(c)). 

To sum up, we have experimentally and theoretically investigated a doubly-driven 1D quasiperiodic lattice that maps to an anisotropic Hofstadter model illuminated by light of tunable polarization. Measurement of the phase diagram demonstrates that the amplitude and polarization of applied radiation coherently control the localization quantum phase transition in 1D~\cite{Lignier_DL_PhysRevLett.99.220403, Shimasaki_PhysRevLett.133.083405, Dotti_doubleLoc_PhysRevResearch.7.L022026}, a phenomenon which maps to switching the direction of conductivity by intense irradiation in the 2D Hofstadter model. 
These results can be interpreted as observation in a 1D quantum system of a driven 2D Hamiltonian important in condensed matter, or equivalently as introducing the concept of variable light polarization to 1D. We observe that linearly and elliptically polarized drives generate strikingly different dynamics for identical drive intensities. We experimentally and theoretically demonstrate that elliptically polarized illumination can generate without any fine-tuned parameters an anomalous critical phase featuring multifractal eigenstates and subdiffusive transport, and confirm experimentally that transport in the anomalous phase is enhanced, rather than suppressed, by the strength of quasiperiodicity.
\yb{
Our protocol challenges the traditional understanding of Floquet multifractality in which localized and ballistic phases are mixed with a low-frequency drive \cite{shimasaki2022anomalous, Gonifmmode_QP_ergodicity_PhysRevB.108.104201}, and where the multifractality is predicted to disappear in the thermodynamic limit. In contrast, our protocol is in the high-frequency regime without any spectral mixing. 
}

\yb{
These results provide a novel and straightforward roadmap for engineering multifractal matter on demand, potentially implementable in a variety of other platforms. While interactions are not important in the dynamics we discuss here, straightforward extensions of the approach could study the same physics in a strongly-interacting regime, probing the existence and stability of a Floquet many-body critical phase~\cite{Wang_MBC_PhysRevLett.126.080602} \rr{or quasi-fractal charge density wave \cite{Gonçalves_2024NatPhys}}.} 
Extensions of techniques demonstrated here could also be used to realize chiral control of magnetism~\cite{Chandran_AAIsing_PhysRevX.7.031061}, 
and probe \yb{higher-dimensional} quantum geometry using \yb{lower-dimensional} circular dichroism~\cite{Ozawa_QGspectroscopy1_PhysRevB.97.201117, Ozawa_QGspectroscopy2_PhysRevResearch.1.032019, Lohse2018_4DQH}.
Driving with more exotic electric fields, for example with Lissajous or unpolarized character, could create novel topological properties \cite{Castro_floquetOCT_PhysRevResearch.4.033213, Mukherjee_FloquetUnpolarized_PhysRevB.98.235112, Quito_unpolarized2_PhysRevLett.126.177201}. The demonstration of spatially inhomogeneous effective tunneling from a global modulation may open up new and simpler avenues to the study of synthetic event horizons~\cite{moghaddam2025synthetichorizonsthermalizationatomic, benhemou2025probingquantumpropertiesblack}.
A naturally complementary direction of investigation is towards solid-state realizations of related light-induced phenomena. The approach presented here can provide both path-finding and motivation for such work and a complementary toolset for joint investigation of nonequilibrium quantum matter \rr{\cite{Lohse2018_4DQH, Paul_PhysRevLett.132.246402, Arlinghaus_Holthaus_2010_OLstrongfieldSim}}. 

\begin{acknowledgments}
We acknowledge experimental assistance from Petros Kousis and Peter Dotti and helpful discussions with Xiao Chai and Jeremy Tanlimco. 
We acknowledge research support from the Air Force Office of Scientific Research (FA9550-20-1-0240), the NSF QLCI program (OMA-2016245), the Noyce Foundation, and the UC Santa Barbara NSF Quantum Foundry funded via the Q-AMASE-i program under Grant DMR1906325. A.R.D. acknowledges support from the NSF NRT program under grant 2152201. The development of the bichromatic optical lattice techniques which are the basis for this work was supported by the U.S. Department of Energy, Office of Science, National Quantum Information Science Research Centers, Quantum Science Center.

Y.B., A.R.D. and T.S. prepared the experimental setup and performed the experiments. Y.B. conceptualized the project and performed theoretical calculations and data analysis. D.M.W. supervised the project. All authors contributed substantially to the work presented in this paper, including discussions of data and preparation of the manuscript.  All authors declare no competing interests. 

\end{acknowledgments}

\appendix

\section{Methods}\label{apd:methods}
The experiments described in the main text commence by adiabatically loading a Bose-Einstein condensate of about $2\times 10^{5}$ $^{84}$Sr atoms into the combination of a 1D bichromatic optical lattice along an axis perpendicular to gravity and a weak crossed dipole trap. The experiments begin when the crossed dipole trap is suddenly switched off and the lattice modulation simultaneously starts.  
 
The primary lattice is generated by two counterpropagating laser beams with wavelength $\lambda_P= 1064\,\mathrm{nm}$. One of the beams composing the primary lattice has a greater intensity to provide transverse confinement and support the atoms against gravity. The combined potential including gravity  has a trap depth of about $1.4\,\mu \mathrm{K}$. We estimate the radial and longitudinal trap frequencies of to be $\omega_r= 2\pi\times 117.8 \, \mathrm{Hz} $ and $\omega_\mathrm{long} = 2\pi\times 0.36 \, \mathrm{Hz}$. 
The amplitudes and frequencies of the two beams comprising the primary lattice are independently controlled using acousto-optic modulators (AOMs). In all experiments, $V_P$ is maintained at $9E_{\mathrm{rec},P}$, where $E_{\mathrm{rec},P}=\hbar^2 k^2_P/2m$, calibrated first with Kapitza-Dirac diffraction and then with dipolar modulation spectroscopy \cite{Tokuno_dipSpec_PhysRevLett.106.205301}. 
    
The two laser beams that compose the secondary lattice have wavelength $\lambda_S=874.61 \, \mathrm{nm}$ and are derived from a continuous-wave titanium-sapphire laser whose wavelength is stabilized against drift using a wavemeter. The two beams have roughly equal intensity and are adjusted in tandem to control the depth of the secondary lattice $V_S$. Similarly to the case of the primary lattice, the amplitude and frequency of each secondary lattice beam is independently controlled using AOMs.

\rr{Expansion of the atoms is quantified by fitting the density distribution to a narrow Gaussian atop a broad shallow pedestal which due to a background of thermally excited atoms. }

The radiofrequency drives of the AOMs employed in the position modulation of the lattices are generated by 2-channel arbitrary waveform generators (Keysight 33600A) such that the first channel modulates the frequency of the second channel which is applied to the AOMs. All sources share a clock signal derived from the same Rubidium reference (FS725, Stanford Research Systems). 

\begin{figure*}[ht!]
    \centering
    \includegraphics[width=180mm]{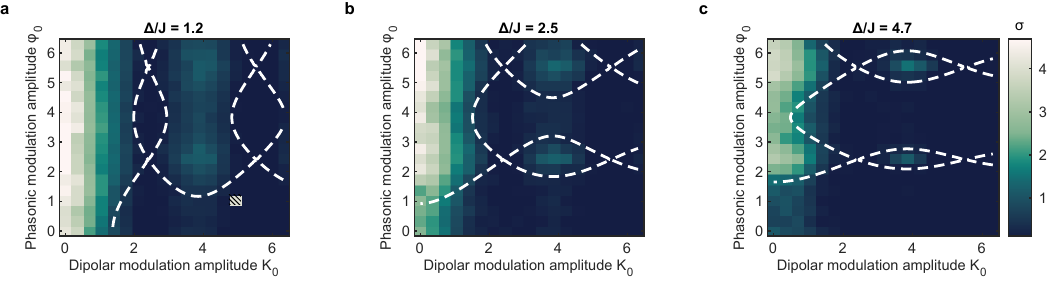}
    \caption{\textbf{Unscaled fractional expansion $\sigma$ under dipolar and phasonic amplitudes.} While the delocalized regions are still visible, their contrast is lower due to effectively reduced kinetic energy. Dashed white curves represent the theoretically predicted phase boundary. }
    \label{fig_ED:unscaledPD}
\end{figure*}
\section{Phase diagrams without the Bessel function scaling}
We report the drive-controlled phase diagram probed by an effective width $\sigma_\mathrm{eff}$, defined as the fractional expansion scaled by the \rr{norm of the} zeroth Bessel function of the dipolar amplitude, $\vert\mathcal{J}_0 (K_0)\vert$. This is motivated by the fact that the expansion speed is proportional to the bandwidth in a clean lattice. \yb{
In Figure \ref{fig_ED:unscaledPD},} we include the unscaled phase diagrams with the fractional expansion. While the contrast is lower, these phase diagrams still clearly demonstrates the phase boundary. 

\section{Anomalous expansion dependence on driving frequency}
\begin{figure}[h!]
    \centering
    \includegraphics[width= 89mm]{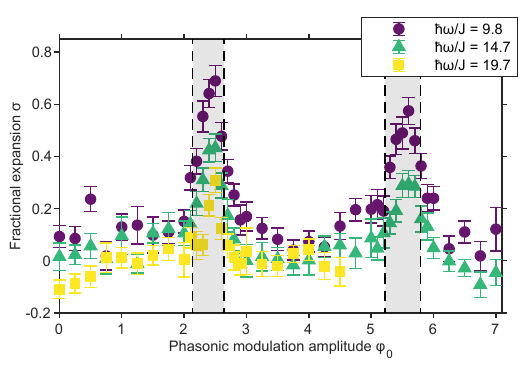}
    \caption{\textbf{Dependence of anomalous expansion under elliptically polarized modulation with respect to the driving frequency $\hbar\omega$}. The expansion is suppressed as the driving frequency is increased, according to Eq. \ref{eq:j1}. The condensate expands for $t_\mathrm{hold} = 10\, \mathrm{s}$ for all cases here. Only part of the data for $\hbar\omega =19.7J$ are displayed due to excessive atom loss from large phasonic amplitude and high driving frequency. Error bars are one s.e.m.}
    \label{fig_ED:hw dependence}
\end{figure}

The tunneling strength $J_\mathrm{EP}$ in Eq.(\ref{eq:j1}), characterizing the strength of the anomalous expansion, decreases as the driving frequency increases. Thus the anomalous expansion should be suppressed with higher driving frequency. Results shown in Fig.\ref{fig_ED:hw dependence} directly test this prediction. Indeed, we observed a slower anomalous expansion with increasing driving frequency. These results provide another support of our theoretical modeling through effective Hamiltonian. 

\begin{figure}[h!]
    \centering
    \includegraphics[width = 89mm]{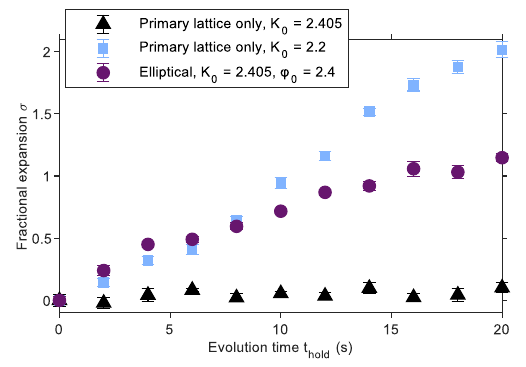}
    \caption{
        \yb{
            \textbf{Benchmarking condensate expansion}. Measured width of the condensate during time evolution of up to $t_\mathrm{hold} = 20 \, \mathrm{s}$. For the case with primary lattice only, ballistic expansion and dynamic localization are maintained during the entirety of the time evolution. The case with critical sub-diffusive expansion is also shown for comparison. Errorbars are one s.e.m.}
    }
    \label{figSI:threeCasesBenchmark}
\end{figure}

\yb{
\section{Experimental benchmark of the evolution}\label{apd:benchmark}
We experimentally benchmark our expansion dynamics up to the longest possible evolution time of $t_\mathrm{hold} = 20\,\mathrm{s} \approx 6400 T_J$. Specifically, we measure the expansion of the condensate in the shaken primary lattice only (that is, $V_S = 0$), at $K_0 = 2.2$ and $K_0 = K_0^\mathrm{DL} = 2.405$. Since there is no quasiperiodicity, the condensate expands ballistically when $K_0 = 2.2$ despite a greatly reduced kinetic energy ($\mathcal{J}_0(K_0) \approx 0.11$), and dynamically localizes when $K_0 = K_0^\mathrm{DL}$. The resulting data are shown in Fig. \ref{figSI:threeCasesBenchmark}, demonstrating that we observed the expected ballistic and dynamically localized dynamics for up to $t_\mathrm{hold} = 20\,\mathrm{s}$, indicating that interaction effects did not qualitatively affect either state.  We further include the measured expansion close to the first peak of the anomalous phase ($K_0 = 2.405$, $\varphi_0 = 2.4$, $\vartheta = \pi/2$), which shows the characteristic sub-ballistic expansion. Since its width is smaller than its ballistic counterpart during the time evolution, we conclude that the reduced expansion is not due to other experimental imperfections such as residual longitudinal trapping but rather reflects the physics discussed in the main text. 
}

\yb{
\section{A short review of the Harper-Hofstatder mapping} \label{apd:HHmap}
Here we briefly review the cross-dimensional mapping from the 2D Harper-Hofstadter model to the 1D Aubry-Andr\'e-Harper model, first described by Harper \cite{harper_single_1955} and
subsequently Hofstadter \cite{Hofstadter_OG_PhysRevB.14.2239} and others.
The 2D Harper-Hofstatder model describes a non-interacting electron in a 2D rectangular lattice, with nearest-neighbor hoppping and a magnetic flux $\beta = \Phi/\Phi_0$ per plaquette, where $\Phi_0$ is the flux quantum. In the Landau gauge, the Hamiltonian is: 
\begin{equation}
    \hat H_\mathrm{Harper} = \sum_{j,l} -J \hat d_{j+1, l}^\dagger \hat d_{j,l} + 
    J_y e^{-i 2\pi\beta j} \hat d_{j, l+1}^\dagger \hat d_{j,l} + \mathrm{h.c.},  \label{eqS:harper}
\end{equation}
where $J$ is the tunneling strength along the $x$-direction (site labelled by $j$) and $J_y := \Delta/2$ the tunneling strength along the $y$-direction (labelled by $l$), and $d_{j,l}^{(\dagger)}$ is the annihilation (creation) operator in 2D. 
This 2D Hamiltonian is independent of the $y$-direction, so we can employ a plane wave basis along that direction. We define the annihilation (creation) operators $\hat c_{j, \kappa}^{(\dagger)}$ with anticommutation relation $\{\hat c_{j, \kappa}, \hat c_{j', \kappa'}^{\dagger}\} = \delta_{j,j'} \delta_{\kappa, \kappa'}$. They are related to the 2D operators as $\hat c_{j, \kappa}^{(\dagger)} = \sum_l e^{-i \kappa l}\hat d_{j,l}^{(\dagger)}$. With these newly defined operators, we arrive at the 1D Hamiltonian $ H_{\mathrm{AAH}}$:
\begin{align}
    & \hat H_{\mathrm{Harper}} = \int^{2\pi}_0 \frac{\mathrm{d}\kappa}{2\pi} \hat H_{\mathrm{AAH}} (\kappa), \nonumber \\
    & \hat H_{\mathrm{AAH}} (\kappa) = \sum_j -J \hat c_{j+1}^\dagger \hat c_j + \mathrm{h.c.} + \Delta  \cos(2\pi\beta j + \kappa) \hat n_j, 
\end{align}
where $\hat n_j = \hat c_j^\dagger \hat c_j$ and we drop the $\kappa$ dependence in the operator. Accordingly, we refer to the $x$-direction as the ``physical'' dimension and the $y$-direction as the  ``additional'' dimension. 
}

\yb{
This 1D Hamiltonian is known as the Aubry-Andr\'e-Harper (AAH) model \cite{aubry1980analyticity} and can be thought of as representing a 1D lattice with a spatially modulated chemical potential. Here $\Delta$ becomes the strength of the modulating potential and $\beta$, the incommensurate ratio, is the ratio between the period of the potential modulation and the lattice spacing. When $\beta$ is irrational the system is quasiperiodic and exhibits a localization quantum phase transition from a delocalized ballistic phase ($\Delta<2J$) to a localized phase ($\Delta >2J$) with exponentially localized wave functions in position space. 
}

\yb{
The inclusion of external driving field $\mathbf{A}(t)$ is achieved by essentially the same procedure with the addition of Peierls substitution, replacing $J_{x,y} \rightarrow J_{x,y} \exp{(-i A_{x,y}(t))}$ respectively. In the Supplementary Material \cite{supp_mat}, we discuss the dynamical consequence of this mapping and its inclusion of the driving field in details. 
}

\section{From shaken 1D bichromatic lattice to doubly driven AAH model}
In this section, we detail the recipe to create the polarized driving protocol through combined sublattice modulation. 

For ultracold atoms trapped in a doubly-driven 1D bichromatic lattice potential, the Hamiltonian is
\begin{multline}
    \hat H_\mathrm{bic} = \frac{\hat p^2}{2m} + \frac{V_P}{2}\cos \left(2 k_P (\hat x-x_P(t)) \right) \\
    + \frac{V_S}{2} \cos \left(2 k_S (\hat x-x_S(t)) + \kappa \right), \label{eq:bichromaticLatt}
\end{multline}
where $m$ is the mass of the trapped atoms, $k_{P,S}$ are the wave vectors and $V_{P,S}$ the depths of the primary and secondary lattice, respectively. The time-dependent terms $x_{P,S}(t)$ describe the position modulation (``shaking") of the primary and secondary lattice, respectively, which are independently controlled. Only sinusoidal modulation is used in this work, so $x_{P,S} (t) = \alpha_{P,S} \sin (\omega t + \theta_{P,S})$ with amplitudes $\alpha_{P,S}$, phases $\theta_{P,S}$ and identical driving frequency $\omega$.
A unitary transformation to the co-moving frame of the moving primary lattice shows that the inertial force from $x_P(t)$ acts as a dipolar, spatially homogeneous AC field $F(t) := m\ddot x_P (t)$:
\begin{multline}
    \hat H_\mathrm{bic} \rightarrow \frac{\hat p^2}{2m} + V_P \cos \big ( 2k_P \hat x \big) \\
    + V_S \cos \big(2k_S (\hat x - x_\mathrm{phason}(t) ) + \kappa \big) + \hat x F(t). \label{eq:expHamil}
\end{multline}

In this frame, the modulation of the secondary lattice, $x_\mathrm{phason}(t)$, depends on the interference between $x_P(t)$ and $x_S(t)$: 
\begin{align}
    & x_\mathrm{phason} = x_S(t)-x_P (t) := \alpha_\mathrm{phason}\sin(\omega t + \theta_\mathrm{phason}), \\
    & \alpha_\mathrm{phason} = \pm \sqrt{\alpha_P^2 + \alpha_S^2 - 2 \alpha_P \alpha_S \cos(\theta_S - \theta_P)} \\
    & \theta_\mathrm{phason} = \operatorname{atan2}(u,v), \label{eq:phasonparams}
\end{align}
where we define $u := {\alpha_S \sin \theta_S - \alpha_P \sin \theta_P}$, $v := \alpha_S \cos \theta_S - \alpha_P \cos \theta_P$ and $\operatorname{atan2}(u,v)$ is the 2-argument arctangent function.

We keep $V_P\gg V_S$ such that the locations of potential minima of primary lattice are not significantly altered by the secondary lattice \cite{Modugno_2009}. We further restrict the driving frequency to be within the first gap of the unperturbed primary lattice. Under these conditions, the single-band, tight-binding approximation applies and the proper Hamiltonian can be written as $\hat H = \sum_j \hat H_j$ where
\begin{align}
    \hat H_j = &-J \hat c_{j+1}^\dagger \hat c_j + \mathrm{h.c.} \nonumber \\
    &+ \Delta \cos(2\pi\beta j + \varphi_0 \sin(\omega t + \theta_\mathrm{phason}) + \kappa) \hat n_j \nonumber\\
    &+ j K\cos(\omega t + \theta_x) \hat n_j
\end{align}
where $\hat c_j^{(\dagger)}$ is the annihilation (creation) operator at site $j$, $\hat n_j$ is the density operator at site $j$ and $K$ is the strength of the force. Since the external force exactly resembles the dipolar force on a charge exposed to a spatially homogeneous, linear field, it is referred to as the dipolar modulation. The phasonic modulation $\varphi_0 \sin\omega t$ refers to the time-dependence of the phason in the incommensurate potential. A further frame transformation to the velocity gauge absorbs the linear oscillating force into an oscillating phase in the tunneling term: 
\begin{align}
    \hat H_j = &-J e^{-i K_0 \sin(\omega t + \theta_x)} \hat c_{j+1}^\dagger \hat c_j + \mathrm{h.c.} \nonumber \\ 
    &+ \Delta \cos(2\pi\beta j + \kappa +  \varphi_0\sin(\omega t + \theta_\mathrm{phason} )) \hat n_j, \label{DAAH2}
\end{align}
We thus recover the doubly-driven AAH model from the experimentally relevant Hamiltonian Eq.(\ref{eq:expHamil}). 

To connect this Hamiltonian to what we actually control in the experiment, we first review and describe frequency modulation of an optical lattice, often referred to as ``shaking", a well-established technique for generating artificial gauge fields. We then extend such modulation to a shaken bichromatic lattice and connect the experimental parameters to those in the DDAAH model. Our method enables ``polarized" dipolar and phasonic driving through time-dependent spatial translations $x_{P(S)}(t)$ of each of the optical lattice potentials along the lattice direction.
    We employ sinusoidal frequency modulation on the primary and secondary lattices:
    \begin{equation}
        f_{P} (t) =f_{0, P}\sin(\omega t), \quad f_S (t) = f_{0,S}\sin(\omega t +\phi),
    \end{equation}
where $f_{0,P(S)}$ is the experimentally relevant amplitude of frequency deviation between the laser beams forming the primary (secondary) lattice and sets the amplitude of the frequency modulation waveform, $\omega$ is the angular modulation frequency, and $\phi$ is the phase difference between the modulation waveforms on the primary and secondary lattices. In particular, we note that the phase difference $\phi$ does not equate to the polarization phase $\vartheta$. Their relation will be clarified shortly. The range of dimensionless shaking amplitudes $K_0$ and $\varphi_0$ examined in the experiments described in this paper is thus realized through appropriately set amplitudes and phases of the two frequency modulation waveforms $f_{0,P}$ and $f_{0,S}$.
    
The sinusoidal frequency modulation leads to time dependent position modulations
\begin{align}
    & x_P(t)=\frac{d_P f_{0,P}}{\omega}- \frac{d_P f_{0,P}}{\omega}\mathrm{cos}(\omega t)\\
    & x_S(t)=\frac{d_S f_{0,S}}{\omega}\mathrm{cos}(\phi)- \frac{d_S f_{0,S}}{\omega}\mathrm{cos}(\omega t + \phi)
\end{align}
    where $d_{P(S)}=\lambda_{P(S)}/2$ is the lattice spacing of the primary (secondary) lattice. Rewriting the position modulations as sinusoidal modulations $x_{P(S)}=\alpha_{P(S)}\mathrm{sin}(\omega t+ \theta_{P(S)})$ leads to the expression of the shaking amplitudes $\alpha_{P(S)}$ and phases $\theta_{P(S)}$ in terms of the controllable parameters of the frequency modulation waveform:
\begin{equation}
    \alpha_{P(S)}=\frac{d_{P(S)}f_{0,P(S)}}{\omega}, \quad \theta_{P}=-\frac{\pi}{2}, \quad \theta_S=\phi-\frac{\pi}{2}.
\end{equation}

We now relate the experimentally accessible parameters to the parameters in the DDAAH Hamiltonian. The dimensionless dipolar shaking amplitude $K_0$, which characterizes the amplitude of the field component parallel to the physical dimension, is determined only by the shaking amplitude $\alpha_P$ of the primary lattice:
\begin{equation}
    K_0 = \frac{\pi}{\hbar k_P}m\omega\alpha_P = \frac{\pi}{4}\frac{f_{0,P}}{f_{\mathrm{rec},P}},
\end{equation}
where $f_{\mathrm{rec},P}$ is the recoil frequency of the primary lattice. 
The phasonic modulation $x_{\mathrm{phason}}(t)$ is the time-dependent difference in position of the secondary and primary lattices, $x_{\mathrm{phason}}(t) = x_{S}(t)-x_{P}(t)$. Thus, $\alpha_{\mathrm{phason}}$ depends on both amplitudes $\alpha_P$ and $\alpha_S$ as well as the phase difference between the modulation waveforms as expressed in Eq. (\ref{eq:phasonparams}). The dimensionless phasonic shaking amplitude $\varphi_0$ and the phase difference $\theta_{\mathrm{phason}}$ are 
\begin{align}   
    \varphi_0 &= 2k_s\alpha_{\mathrm{phason}} \nonumber \\
    &= \pm \frac{\beta}{f}\sqrt{f^2_{0,P}+\frac{1}{\beta^2}f^2_{0,S}-\frac{2}{\beta}f_{0,P}f_{0,S}\mathrm{cos}(\phi)}, \\
    \theta_{\mathrm{phason}} &= \mathrm{atan2}(\beta f_{0,P}-f_{0,S}\mathrm{cos}\phi, f_{0,S}\mathrm{sin}\phi)
\end{align}
in terms of the modulation waveform parameters, where $\beta=\lambda_P/\lambda_S$ again is the incommensurate ratio of the bichromatic lattice 
\yb{and $\mathrm{atan2}(y,x)$ is the two-argument arctangent}. We find that choosing the expression of $\varphi_0$ with the negative sign offers us greater experimental tunability; this choice simply corresponds to a convention regarding the chirality of the polarization. 

Finally, the polarization phases $\theta_x, \theta_y$ and $\vartheta = \theta_y-\theta_x$ are simply
\[
\theta_x = \theta_P + \frac{\pi}{2} =0, \quad \theta_y = \theta_\mathrm{phason} \longrightarrow \vartheta = \theta_\mathrm{phason}.
\]

All relevant relations between the driving parameters in the Hamiltonian (Eq. \ref{eq:DDAAH_comoving}) and the experimentally controlled quantities are summarized in Table~\ref{tabley}.
\renewcommand{\arraystretch}{1}
\begin{table}[h!]
\begin{ruledtabular}
\begin{tabular}{ c c }

\textbf{Parameters} & \textbf{Expressions}                                                                                             \\ \hline
Dipolar amplitude $K_0$               & $({\pi}/{4}){f_{0,P}}/{f_{\mathrm{rec},P}}$                                                                \\ 
Phasonic amplitude $\varphi_0$         & $\pm \frac{\beta}{f}\sqrt{f^2_{0,P}+\frac{1}{\beta^2}f^2_{0,S}-\frac{2}{\beta}f_{0,P}f_{0,S}\mathrm{cos}(\phi)}$ \\ 
Polarization $\vartheta$         & $\mathrm{atan2}(\beta f_{0,P}-f_{0,S}\mathrm{cos}\phi, f_{0,S}\mathrm{sin}\phi)$                                 \\ 
\end{tabular}
\caption{Relations of drive parameters in the Hamiltonian (Eq.\ref{eq:DDAAH_comoving}) to directly experimentally controllable quantities.}
\label{tabley}
\end{ruledtabular}
\end{table}

We now discuss how the polarization of the drive is controlled in the experiment. Operationally, we first choose the desired dimensionless parameters $K_0$, $\varphi_0$ and the polarization $\vartheta$, and then compute the required experimental parameters $f_{0,P}$, $f_{0,S}$ and $\phi$ which are sent to the AWG. Thus, in this section, we express these experimental parameters in terms of the dimensionless model parameters $K_0$, $\varphi_0$ and $\vartheta$.
Here we explicitly restrict the term ``elliptical polarization'' to the case of $\vartheta =\pm \pi/2$ (with $K_0, \varphi_0 \neq 0$). That is, we restrict ourselves to the case where the axes of the polarization ellipse of the light are aligned to the lattice axes. Other cases of elliptical polarization are outside the scope of this work but can be achieved by generalizing the recipe given here. 

The condition $\vartheta =\pm \pi/2$ is most simply accomplished experimentally by enforcing the condition
\begin{equation}
    f_{0,S}\sin\phi = 0. \label{eq:ellPolCondition}
\end{equation}
Either $f_{0,S} = 0$ \cite{Shimasaki_PhysRevLett.133.083405} or $\phi = 0, \; \pm \pi$ would satisfy this condition. We found that choosing the two modulations to be in phase ($\phi=0$) offers the greatest tunability of the phasonic modulation amplitude $\varphi_0$. Under this condition, the position modulations $x_P(t)$ and $x_S(t)$ destructively interfere and the phasonic amplitude is given by
\begin{equation}
    \varphi_0 = -\frac{1}{f}(\beta f_{0,P} - f_{0,S}) = \frac{4\beta}{\pi} \frac{f_\mathrm{rec}}{f}K_0 - \frac{f_{0,S}}{f}. \label{eq:phasonAmpEP}
\end{equation}
This gives the desired frequency modulation amplitude for the secondary lattice as
\begin{equation}
    f_{0,S} = f \varphi_0 + \beta f_{0,P}.
\end{equation}

For linearly polarized drive ($\vartheta =0,\pm\pi$), we require that
\begin{equation}
    \beta f_{0,P}=f_{0,S}\cos\phi \quad \mathrm{and} \quad f_{0,S} \sin\phi \neq 0. \label{eq:LPcondition}
\end{equation}
The first equality in the above condition implies that
\[
f\varphi_{0} = f_{0,S}\sin\phi.
\]
We restrict $\phi \in [0,\pi/4]$ such that $\varphi_0 \geq 0$. The condition of linear polarization can then be re-expressed as
\begin{equation}
    f_{0,S}\cos\phi = \beta f_{0,P}, \quad \mathrm{and} \quad f_{0,S}\sin\phi = f\varphi_0. \label{eq:phasonAmpLP}
\end{equation}
Thus, under the condition of linear polarization, the frequency modulation amplitude of the secondary lattice $f_{0,S}$ and the relative phase between the lattice modulation $\phi$ trace out a circle, in the space of $f_{0,P}$ and $\varphi_0$. 

This geometric picture provides a simple recipe for realizing a linearly polarized drive. We choose the desired values of the dipolar amplitude controlled only by $f_{0,P}$, the phasonic amplitude $\varphi_0$, the driving frequency $f$. Then to achieve a linearly polarized drive, we set
\begin{align}
    & f_{0,S} = \sqrt{ (\beta f_{0,P})^2 + (f\varphi_0)^2}, \\
    & \phi = \omega \tau_\mathrm{delay} = \arctan \left(\frac{f\varphi_0}{\beta f_{0,P}}\right) = \arctan \left( \frac{\pi}{4\beta} \frac{f}{f_\mathrm{rec,P}} \frac{\varphi_0}{K_0}\right).
\end{align}
An experiment using linearly polarized modulation begins first with modulation of the primary lattice only. The relative phase $\phi$ between the two lattice modulations  is controlled by an initial delay of duration $\tau_\mathrm{delay}$, after which the modulation of the secondary lattice begins. This delay means that the electric field would first be elliptically polarized because only the primary lattice is shaking ($f_{0,S} (t<\tau_d) = 0$) until $\tau_\mathrm{delay} < 0.25/f$, after which the field maintains linear polarization until the end of modulation. 
Table~\ref{tabley2} summarizes the expressions used to control the drive polarization used in this work. We emphasize that these are not the only possible approaches; a desired polarization is achieved as long as the corresponding conditions (Eq. \ref{eq:ellPolCondition} or Eq. \ref{eq:LPcondition}) are satisfied. 

\renewcommand{\arraystretch}{1}
\begin{table}[h!]
\begin{ruledtabular}
    \begin{tabular}{ccc}
        {\textbf{Parameters}} & \multicolumn{1}{c} {\textit{Elliptical}} & \textit{Linear}                                     \\ \hline
        $f_{0,P}$                                         & \multicolumn{1}{c}{$4 K_0 f_{\mathrm{rec}, P}/\pi$} & {$4 K_0 f_{\mathrm{rec}, P}/\pi$}          \\ 
        $f_{0,S}$                                         & \multicolumn{1}{c}{$ f \varphi_0 + \beta f_{0,P} $} & $\sqrt{ (\beta f_{0,P})^2 + (f\varphi_0)^2}$        \\ 
        $\tau_\mathrm{delay}$                             & \multicolumn{1}{c}{$0$}                                   & $\arctan \left({f\varphi_0}/{\beta f_{0,P}}\right)/\omega$ \\ 
    \end{tabular}
    \caption{Expressions of the experimental parameters for the polarized modulations used in the experiments.}
    \label{tb:experimentalParameters}
    \label{tabley2}
\end{ruledtabular}
\end{table} 

\bibliography{mainBib}

@article{RevModPhys.93.041002,
  title = {Colloquium: Nonthermal pathways to ultrafast control in quantum materials},
  author = {de la Torre, Alberto and Kennes, Dante M. and Claassen, Martin and Gerber, Simon and McIver, James W. and Sentef, Michael A.},
  journal = {Rev. Mod. Phys.},
  volume = {93},
  issue = {4},
  pages = {041002},
  numpages = {31},
  year = {2021},
  month = {Oct},
  publisher = {American Physical Society},
  doi = {10.1103/RevModPhys.93.041002},
  url = {https://link.aps.org/doi/10.1103/RevModPhys.93.041002}
}

@article{Jagannathan,
  title = {Missing link between the two-dimensional quantum Hall problem and one-dimensional quasicrystals},
  author = {Jagannathan, Anuradha},
  journal = {Phys. Rev. B},
  volume = {112},
  issue = {10},
  pages = {L100102},
  numpages = {6},
  year = {2025},
  month = {Sep},
  publisher = {American Physical Society},
  doi = {10.1103/stk9-d9vf},
  url = {https://link.aps.org/doi/10.1103/stk9-d9vf}
}

@article{aubry1980analyticity,
  title={Analyticity breaking and {A}nderson localization in incommensurate lattices},
  author={Aubry, Serge and Andr{\'e}, Gilles},
  journal={Ann. Israel Phys. Soc.},
  volume={3},
  number={133},
  pages={18},
  year={1980}
}

@article{harper_single_1955,
	title = {Single {Band} {Motion} of {Conduction} {Electrons} in a {Uniform} {Magnetic} {Field}},
	volume = {68},
	issn = {0370-1298},
	url = {https://iopscience.iop.org/article/10.1088/0370-1298/68/10/304},
	doi = {10.1088/0370-1298/68/10/304},
	number = {10},
	urldate = {2021-08-12},
	journal = {Proceedings of the Physical Society. Section A},
	author = {Harper, P G},
	month = oct,
	year = {1955},
	pages = {874--878},
}

@article{Hofstadter_OG_PhysRevB.14.2239,
  title = {Energy levels and wave functions of Bloch electrons in rational and irrational magnetic fields},
  author = {Hofstadter, Douglas R.},
  journal = {Phys. Rev. B},
  volume = {14},
  issue = {6},
  pages = {2239--2249},
  numpages = {0},
  year = {1976},
  month = {Sep},
  publisher = {American Physical Society},
  doi = {10.1103/PhysRevB.14.2239},
  url = {https://link.aps.org/doi/10.1103/PhysRevB.14.2239}
}

@Article{Lohse2018_4DQH,
author={Lohse, Michael
and Schweizer, Christian
and Price, Hannah M.
and Zilberberg, Oded
and Bloch, Immanuel},
title={Exploring 4D quantum Hall physics with a 2D topological charge pump},
journal={Nature},
year={2018},
month={Jan},
day={01},
volume={553},
number={7686},
pages={55-58},
issn={1476-4687},
doi={10.1038/nature25000},
url={https://doi.org/10.1038/nature25000}
}

@article{Chandran_AAIsing_PhysRevX.7.031061,
  title = {Localization and Symmetry Breaking in the Quantum Quasiperiodic Ising Glass},
  author = {Chandran, A. and Laumann, C. R.},
  journal = {Phys. Rev. X},
  volume = {7},
  issue = {3},
  pages = {031061},
  numpages = {21},
  year = {2017},
  month = {Sep},
  publisher = {American Physical Society},
  doi = {10.1103/PhysRevX.7.031061},
  url = {https://link.aps.org/doi/10.1103/PhysRevX.7.031061}
}

@misc{benhemou2025probingquantumpropertiesblack,
      title={Probing quantum properties of black holes with a Floquet-driven optical lattice simulator}, 
      author={Asmae Benhemou and Georgia Nixon and Aydin Deger and Ulrich Schneider and Jiannis K. Pachos},
      year={2025},
      eprint={2312.14058},
      archivePrefix={arXiv},
      primaryClass={cond-mat.quant-gas},
      url={https://arxiv.org/abs/2312.14058}, 
}

@misc{moghaddam2025synthetichorizonsthermalizationatomic,
      title={Synthetic Horizons and Thermalization in an Atomic Chain and its Relation to Quantum Hall Systems}, 
      author={Ali G. Moghaddam and Viktor Könye and Lotte Mertens and Jasper van Wezel and Jeroen van den Brink},
      year={2025},
      eprint={2504.16194},
      archivePrefix={arXiv},
      primaryClass={cond-mat.mes-hall},
      url={https://arxiv.org/abs/2504.16194}, 
}

@article{Barelli_PhysRevLett.83.5082,
  title = {Magnetic-Field-Induced Directional Localization in a 2D Rectangular Lattice},
  author = {Barelli, A. and Bellissard, J. and Claro, F.},
  journal = {Phys. Rev. Lett.},
  volume = {83},
  issue = {24},
  pages = {5082--5085},
  numpages = {0},
  year = {1999},
  month = {Dec},
  publisher = {American Physical Society},
  doi = {10.1103/PhysRevLett.83.5082},
  url = {https://link.aps.org/doi/10.1103/PhysRevLett.83.5082}
}

@article{Paul_PhysRevLett.132.246402,
  title = {Directional Localization from a Magnetic Field in Moir\'e Systems},
  author = {Paul, Nisarga and Crowley, Philip J. D. and Fu, Liang},
  journal = {Phys. Rev. Lett.},
  volume = {132},
  issue = {24},
  pages = {246402},
  numpages = {6},
  year = {2024},
  month = {Jun},
  publisher = {American Physical Society},
  doi = {10.1103/PhysRevLett.132.246402},
  url = {https://link.aps.org/doi/10.1103/PhysRevLett.132.246402}
}

@misc{Paul_Crowley_2024_interaction, 
    title={Stability of sub-dimensional localization to electronic interactions}, 
    author={Paul, Nisarga and Crowley, Philip J. D.},   
    url={http://arxiv.org/abs/2410.21408}, 
    DOI={10.48550/arXiv.2410.21408}, 
    eprint={2410.21408}, 
    archivePrefix={arXiv}, 
    year={2024}, 
    month=oct }

@misc{zhou2025fundamentallocalizationphasesquasiperiodic,
      title={The fundamental localization phases in quasiperiodic systems: A unified framework and exact results}, 
      author={Xin-Chi Zhou and Bing-Chen Yao and Yongjian Wang and Yucheng Wang and Yudong Wei and Qi Zhou and Xiong-Jun Liu},
      year={2025},
      eprint={2503.24380},
      archivePrefix={arXiv},
      primaryClass={cond-mat.dis-nn},
      url={https://arxiv.org/abs/2503.24380}, 
}

@misc{huang2025exactquantumcriticalstates,
      title={Exact quantum critical states with a superconducting quantum processor}, 
      author={Wenhui Huang and Xin-Chi Zhou and Libo Zhang and Jiawei Zhang and Yuxuan Zhou and Bing-Chen Yao and Zechen Guo and Peisheng Huang and Qixian Li and Yongqi Liang and Yiting Liu and Jiawei Qiu and Daxiong Sun and Xuandong Sun and Zilin Wang and Changrong Xie and Yuzhe Xiong and Xiaohan Yang and Jiajian Zhang and Zihao Zhang and Ji Chu and Weijie Guo and Ji Jiang and Xiayu Linpeng and Wenhui Ren and Yuefeng Yuan and Jingjing Niu and Ziyu Tao and Song Liu and Youpeng Zhong and Xiong-Jun Liu and Dapeng Yu},
      year={2025},
      eprint={2502.19185},
      archivePrefix={arXiv},
      primaryClass={quant-ph},
      url={https://arxiv.org/abs/2502.19185}, 
}

@article{Itzler_Bojko_Chaikin_1992, title={Anisotropic Localization in Periodic Superconducting Networks}, volume={20}, ISSN={0295-5075, 1286-4854}, DOI={10.1209/0295-5075/20/7/011}, number={7}, journal={Europhysics Letters (EPL)}, author={Itzler, M. A and Bojko, R and Chaikin, P. M}, year={1992}, month=dec, pages={639–644} }

@article{Itzler_Danner_Bojko_Chaikin_1994, title={ac-susceptibility measurements on isotropic and anisotropic superconducting networks}, volume={49}, rights={http://link.aps.org/licenses/aps-default-license}, ISSN={0163-1829, 1095-3795}, DOI={10.1103/PhysRevB.49.6815}, number={10}, journal={Physical Review B}, author={Itzler, M. A. and Danner, G. M. and Bojko, R. and Chaikin, P. M.}, year={1994}, month=mar, pages={6815–6821},  }

@article{Sun_Ralston_1991, title={Duality symmetry and power-law fading of frustration in a quantum multiconnected superconductor}, volume={43}, rights={http://link.aps.org/licenses/aps-default-license}, ISSN={0163-1829, 1095-3795}, DOI={10.1103/PhysRevB.43.5375}, number={7}, journal={Physical Review B}, author={Sun, Sheng Nien and Ralston, John P.}, year={1991}, month=mar, pages={5375–5380}}

@misc{supp_mat,
    note={See Supplementary Material for additional theoretical and experimental details, including more discussions on the Hamiltonians and the numerical methods, a detailed recipe of creating polarized drive with frequency-modulated bichromatic optical lattice, calibration of dipolar and phasonic modulation amplitudes and condensate expansion in a driven lattice. 
         }
}

@article{Shimasaki_PhysRevLett.133.083405,
  title = {Reversible Phasonic Control of a Quantum Phase Transition in a Quasicrystal},
  author = {Shimasaki, Toshihiko and Bai, Yifei and Kondakci, H. Esat and Dotti, Peter and Pagett, Jared E. and Dardia, Anna R. and Prichard, Max and Eckardt, Andr\'e and Weld, David M.},
  journal = {Phys. Rev. Lett.},
  volume = {133},
  issue = {8},
  pages = {083405},
  numpages = {6},
  year = {2024},
  month = {Aug},
  publisher = {American Physical Society},
  doi = {10.1103/PhysRevLett.133.083405},
  url = {https://link.aps.org/doi/10.1103/PhysRevLett.133.083405}
}

@article{Lignier_DL_PhysRevLett.99.220403,
  title = {Dynamical Control of Matter-Wave Tunneling in Periodic Potentials},
  author = {Lignier, H. and Sias, C. and Ciampini, D. and Singh, Y. and Zenesini, A. and Morsch, O. and Arimondo, E.},
  journal = {Phys. Rev. Lett.},
  volume = {99},
  issue = {22},
  pages = {220403},
  numpages = {4},
  year = {2007},
  month = {Nov},
  publisher = {American Physical Society},
  doi = {10.1103/PhysRevLett.99.220403},
  url = {https://link.aps.org/doi/10.1103/PhysRevLett.99.220403}
}

@article{Goldman_PhysRevX.4.031027,
  title = {Periodically Driven Quantum Systems: Effective Hamiltonians and Engineered Gauge Fields},
  author = {Goldman, N. and Dalibard, J.},
  journal = {Phys. Rev. X},
  volume = {4},
  issue = {3},
  pages = {031027},
  numpages = {29},
  year = {2014},
  month = {Aug},
  publisher = {American Physical Society},
  doi = {10.1103/PhysRevX.4.031027},
  url = {https://link.aps.org/doi/10.1103/PhysRevX.4.031027}
}

@article{Eckardt_2015_HFE,
doi = {10.1088/1367-2630/17/9/093039},
url = {https://dx.doi.org/10.1088/1367-2630/17/9/093039},
year = {2015},
month = {sep},
publisher = {IOP Publishing},
volume = {17},
number = {9},
pages = {093039},
author = {André Eckardt and Egidijus Anisimovas},
title = {High-frequency approximation for periodically driven quantum systems from a Floquet-space perspective},
journal = {New Journal of Physics}
}

@article{Tokuno_dipSpec_PhysRevLett.106.205301,
  title = {Spectroscopy for Cold Atom Gases in Periodically Phase-Modulated Optical Lattices},
  author = {Tokuno, Akiyuki and Giamarchi, Thierry},
  journal = {Phys. Rev. Lett.},
  volume = {106},
  issue = {20},
  pages = {205301},
  numpages = {4},
  year = {2011},
  month = {May},
  publisher = {American Physical Society},
  doi = {10.1103/PhysRevLett.106.205301},
  url = {https://link.aps.org/doi/10.1103/PhysRevLett.106.205301}
}

@article{Modugno_2009,
doi = {10.1088/1367-2630/11/3/033023},
url = {https://dx.doi.org/10.1088/1367-2630/11/3/033023},
year = {2009},
month = {mar},
publisher = {},
volume = {11},
number = {3},
pages = {033023},
author = {Modugno, Michele},
title = {Exponential localization in one-dimensional quasi-periodic optical lattices},
journal = {New Journal of Physics}
}

@article{Han_cri_bicrit_modifiedAAH_PhysRevB.50.11365,
  title = {Critical and bicritical properties of Harper's equation with next-nearest-neighbor coupling},
  author = {Han, J. H. and Thouless, D. J. and Hiramoto, H. and Kohmoto, M.},
  journal = {Phys. Rev. B},
  volume = {50},
  issue = {16},
  pages = {11365--11380},
  numpages = {0},
  year = {1994},
  month = {Oct},
  publisher = {American Physical Society},
  doi = {10.1103/PhysRevB.50.11365},
  url = {https://link.aps.org/doi/10.1103/PhysRevB.50.11365}
}

@article{Chang_MFanalysis_modifiedAAH_PhysRevB.55.12971,
  title = {Multifractal properties of the wave functions of the square-lattice tight-binding model with next-nearest-neighbor hopping in a magnetic field},
  author = {Chang, I. and Ikezawa, K. and Kohmoto, M.},
  journal = {Phys. Rev. B},
  volume = {55},
  issue = {19},
  pages = {12971--12975},
  numpages = {0},
  year = {1997},
  month = {May},
  publisher = {American Physical Society},
  doi = {10.1103/PhysRevB.55.12971},
  url = {https://link.aps.org/doi/10.1103/PhysRevB.55.12971}
}

@Article{shimasaki2022anomalous,
    author={Shimasaki, Toshihiko
    and Prichard, Max
    and Kondakci, H. Esat
    and Pagett, Jared E.
    and Bai, Yifei
    and Dotti, Peter
    and Cao, Alec
    and Dardia, Anna R.
    and Lu, Tsung-Cheng
    and Grover, Tarun
    and Weld, David M.},
    title={Anomalous localization in a kicked quasicrystal},
    journal={Nature Physics},
    year={2024},
    month={Mar},
    day={01},
    volume={20},
    number={3},
    pages={409-414},
    issn={1745-2481},
    doi={10.1038/s41567-023-02329-4},
    url={https://doi.org/10.1038/s41567-023-02329-4}
}

@article{Ketzmerick_spreading_PhysRevLett.79.1959,
  title = {What Determines the Spreading of a Wave Packet?},
  author = {Ketzmerick, R. and Kruse, K. and Kraut, S. and Geisel, T.},
  journal = {Phys. Rev. Lett.},
  volume = {79},
  issue = {11},
  pages = {1959--1963},
  numpages = {0},
  year = {1997},
  month = {Sep},
  publisher = {American Physical Society},
  doi = {10.1103/PhysRevLett.79.1959},
  url = {https://link.aps.org/doi/10.1103/PhysRevLett.79.1959}
}

@article{Kitagawa_PhysRevB.84.235108,
  title = {Transport properties of nonequilibrium systems under the application of light: Photoinduced quantum Hall insulators without Landau levels},
  author = {Kitagawa, Takuya and Oka, Takashi and Brataas, Arne and Fu, Liang and Demler, Eugene},
  journal = {Phys. Rev. B},
  volume = {84},
  issue = {23},
  pages = {235108},
  numpages = {13},
  year = {2011},
  month = {Dec},
  publisher = {American Physical Society},
  doi = {10.1103/PhysRevB.84.235108},
  url = {https://link.aps.org/doi/10.1103/PhysRevB.84.235108}
}

@Article{Avila2017_proof_CP,
    author={Avila, A.
    and Jitomirskaya, S.
    and Marx, C. A.},
    title={Spectral theory of extended Harper's model and a question by Erd{\H{o}}s and Szekeres},
    journal={Inventiones mathematicae},
    year={2017},
    month={Oct},
    day={01},
    volume={210},
    number={1},
    pages={283-339},
    issn={1432-1297},
    doi={10.1007/s00222-017-0729-1},
    url={https://doi.org/10.1007/s00222-017-0729-1}
}

@article{Bai_PhysRevB.111.115163,
  title = {Tunably polarized driving light controls the phase diagram of one-dimensional quasicrystals and two-dimensional quantum Hall matter},
  author = {Bai, Yifei and Weld, David M.},
  journal = {Phys. Rev. B},
  volume = {111},
  issue = {11},
  pages = {115163},
  numpages = {14},
  year = {2025},
  month = {Mar},
  publisher = {American Physical Society},
  doi = {10.1103/PhysRevB.111.115163},
  url = {https://link.aps.org/doi/10.1103/PhysRevB.111.115163}
}

@article{Ozawa_QGspectroscopy2_PhysRevResearch.1.032019,
  title = {Probing localization and quantum geometry by spectroscopy},
  author = {Ozawa, Tomoki and Goldman, Nathan},
  journal = {Phys. Rev. Res.},
  volume = {1},
  issue = {3},
  pages = {032019},
  numpages = {6},
  year = {2019},
  month = {Nov},
  publisher = {American Physical Society},
  doi = {10.1103/PhysRevResearch.1.032019},
  url = {https://link.aps.org/doi/10.1103/PhysRevResearch.1.032019}
}

@article{Ozawa_QGspectroscopy1_PhysRevB.97.201117,
  title = {Extracting the quantum metric tensor through periodic driving},
  author = {Ozawa, Tomoki and Goldman, Nathan},
  journal = {Phys. Rev. B},
  volume = {97},
  issue = {20},
  pages = {201117},
  numpages = {6},
  year = {2018},
  month = {May},
  publisher = {American Physical Society},
  doi = {10.1103/PhysRevB.97.201117},
  url = {https://link.aps.org/doi/10.1103/PhysRevB.97.201117}
}

@article{Lucioni2011_subdiffMF, 
title={Observation of Subdiffusion in a Disordered Interacting System}, volume={106}, ISSN={0031-9007, 1079-7114}, DOI={10.1103/PhysRevLett.106.230403}, number={23}, journal={Physical Review Letters}, author={Lucioni, E. and Deissler, B. and Tanzi, L. and Roati, G. and Zaccanti, M. and Modugno, M. and Larcher, M. and Dalfovo, F. and Inguscio, M. and Modugno, G.}, year={2011}, month=jun, pages={230403}
}

@article{Wang_MBC_PhysRevLett.126.080602,
  title = {Many-Body Critical Phase: Extended and Nonthermal},
  author = {Wang, Yucheng and Cheng, Chen and Liu, Xiong-Jun and Yu, Dapeng},
  journal = {Phys. Rev. Lett.},
  volume = {126},
  issue = {8},
  pages = {080602},
  numpages = {6},
  year = {2021},
  month = {Feb},
  publisher = {American Physical Society},
  doi = {10.1103/PhysRevLett.126.080602},
  url = {https://link.aps.org/doi/10.1103/PhysRevLett.126.080602}
}

@article{Gonifmmode_QP_ergodicity_PhysRevB.108.104201,
  title = {Quasiperiodicity hinders ergodic Floquet eigenstates},
  author = {Gon\ifmmode \mbox{\c{c}}\else \c{c}\fi{}alves, Miguel and Ribeiro, Pedro and Khaymovich, Ivan M.},
  journal = {Phys. Rev. B},
  volume = {108},
  issue = {10},
  pages = {104201},
  numpages = {10},
  year = {2023},
  month = {Sep},
  publisher = {American Physical Society},
  doi = {10.1103/PhysRevB.108.104201},
  url = {https://link.aps.org/doi/10.1103/PhysRevB.108.104201}
}

@article{Mukherjee_FloquetUnpolarized_PhysRevB.98.235112,
  title = {Floquet topological transition by unpolarized light},
  author = {Mukherjee, Bhaskar},
  journal = {Phys. Rev. B},
  volume = {98},
  issue = {23},
  pages = {235112},
  numpages = {8},
  year = {2018},
  month = {Dec},
  publisher = {American Physical Society},
  doi = {10.1103/PhysRevB.98.235112},
  url = {https://link.aps.org/doi/10.1103/PhysRevB.98.235112}
}

@article{Quito_unpolarized2_PhysRevLett.126.177201,
  title = {Floquet Engineering Correlated Materials with Unpolarized Light},
  author = {Quito, V. L. and Flint, R.},
  journal = {Phys. Rev. Lett.},
  volume = {126},
  issue = {17},
  pages = {177201},
  numpages = {6},
  year = {2021},
  month = {Apr},
  publisher = {American Physical Society},
  doi = {10.1103/PhysRevLett.126.177201},
  url = {https://link.aps.org/doi/10.1103/PhysRevLett.126.177201}
}

@article{Castro_floquetOCT_PhysRevResearch.4.033213,
  title = {Floquet engineering the band structure of materials with optimal control theory},
  author = {Castro, Alberto and De Giovannini, Umberto and Sato, Shunsuke A. and H\"ubener, Hannes and Rubio, Angel},
  journal = {Phys. Rev. Res.},
  volume = {4},
  issue = {3},
  pages = {033213},
  numpages = {11},
  year = {2022},
  month = {Sep},
  publisher = {American Physical Society},
  doi = {10.1103/PhysRevResearch.4.033213},
  url = {https://link.aps.org/doi/10.1103/PhysRevResearch.4.033213}
}

@article{Han_HarperNNN_PhysRevB.50.11365,
  title = {Critical and bicritical properties of Harper's equation with next-nearest-neighbor coupling},
  author = {Han, J. H. and Thouless, D. J. and Hiramoto, H. and Kohmoto, M.},
  journal = {Phys. Rev. B},
  volume = {50},
  issue = {16},
  pages = {11365--11380},
  numpages = {0},
  year = {1994},
  month = {Oct},
  publisher = {American Physical Society},
  doi = {10.1103/PhysRevB.50.11365},
  url = {https://link.aps.org/doi/10.1103/PhysRevB.50.11365}
}

@article{Sun_optimalFreqWindow_PhysRevResearch.2.013241,
  title = {Optimal frequency window for Floquet engineering in optical lattices},
  author = {Sun, Gaoyong and Eckardt, Andr\'e},
  journal = {Phys. Rev. Res.},
  volume = {2},
  issue = {1},
  pages = {013241},
  numpages = {10},
  year = {2020},
  month = {Mar},
  publisher = {American Physical Society},
  doi = {10.1103/PhysRevResearch.2.013241},
  url = {https://link.aps.org/doi/10.1103/PhysRevResearch.2.013241}
}

@article{Chaudhury_transverseInstability_PhysRevA.91.023624,
  title = {Transverse collisional instabilities of a Bose-Einstein condensate in a driven one-dimensional lattice},
  author = {Choudhury, Sayan and Mueller, Erich J.},
  journal = {Phys. Rev. A},
  volume = {91},
  issue = {2},
  pages = {023624},
  numpages = {7},
  year = {2015},
  month = {Feb},
  publisher = {American Physical Society},
  doi = {10.1103/PhysRevA.91.023624},
  url = {https://link.aps.org/doi/10.1103/PhysRevA.91.023624}
}

@article{Weinberg_multiphotonTransition_PhysRevA.92.043621,
  title = {Multiphoton interband excitations of quantum gases in driven optical lattices},
  author = {Weinberg, M. and \"Olschl\"ager, C. and Str\"ater, C. and Prelle, S. and Eckardt, A. and Sengstock, K. and Simonet, J.},
  journal = {Phys. Rev. A},
  volume = {92},
  issue = {4},
  pages = {043621},
  numpages = {10},
  year = {2015},
  month = {Oct},
  publisher = {American Physical Society},
  doi = {10.1103/PhysRevA.92.043621},
  url = {https://link.aps.org/doi/10.1103/PhysRevA.92.043621}
}

@article{Reitter_heating_PhysRevLett.119.200402,
  title = {Interaction Dependent Heating and Atom Loss in a Periodically Driven Optical Lattice},
  author = {Reitter, Martin and N\"ager, Jakob and Wintersperger, Karen and Str\"ater, Christoph and Bloch, Immanuel and Eckardt, Andr\'e and Schneider, Ulrich},
  journal = {Phys. Rev. Lett.},
  volume = {119},
  issue = {20},
  pages = {200402},
  numpages = {6},
  year = {2017},
  month = {Nov},
  publisher = {American Physical Society},
  doi = {10.1103/PhysRevLett.119.200402},
  url = {https://link.aps.org/doi/10.1103/PhysRevLett.119.200402}
}

@article{kraus_quasiperiodicityTranscend_2016,
	title = {Quasiperiodicity and topology transcend dimensions},
	volume = {12},
	issn = {1745-2473, 1745-2481},
	url = {https://www.nature.com/articles/nphys3784},
	doi = {10.1038/nphys3784},
	  
	number = {7},
	urldate = {2023-08-22},
	journal = {Nature Physics},
	author = {Kraus, Yaacov E. and Zilberberg, Oded},
	month = jul,
	year = {2016},
	pages = {624--626},
}

@article{Basov_Averitt_Hsieh_OnDemand_2017, title={Towards properties on demand in quantum materials}, volume={16}, ISSN={1476-1122, 1476-4660}, DOI={10.1038/nmat5017}, number={11}, journal={Nature Materials}, author={Basov, D. N. and Averitt, R. D. and Hsieh, D.}, year={2017}, month=nov, pages={1077–1088}}

@Article{Liu_LSP_10.21468/SciPostPhys.12.1.027,
	title={{Anomalous mobility edges in one-dimensional quasiperiodic models}},
	author={Tong Liu and Xu Xia and Stefano Longhi and Laurent Sanchez-Palencia},
	journal={SciPost Phys.},
	volume={12},
	pages={027},
	year={2022},
	publisher={SciPost},
	doi={10.21468/SciPostPhys.12.1.027},
	url={https://scipost.org/10.21468/SciPostPhys.12.1.027},
}

@Article{Roy_MFwithoutFineTune_10.21468/SciPostPhys.4.5.025,
	title={{Multifractality without fine-tuning in a Floquet quasiperiodic chain}},
	author={Sthitadhi Roy and Ivan M. Khaymovich and Arnab Das and Roderich Moessner},
	journal={SciPost Phys.},
	volume={4},
	pages={025},
	year={2018},
	publisher={SciPost},
	doi={10.21468/SciPostPhys.4.5.025},
	url={https://scipost.org/10.21468/SciPostPhys.4.5.025},
}

@article{Wang_Raman-Critical_PhysRevLett.125.073204,
  title = {Realization and Detection of Nonergodic Critical Phases in an Optical Raman Lattice},
  author = {Wang, Yucheng and Zhang, Long and Niu, Sen and Yu, Dapeng and Liu, Xiong-Jun},
  journal = {Phys. Rev. Lett.},
  volume = {125},
  issue = {7},
  pages = {073204},
  numpages = {7},
  year = {2020},
  month = {Aug},
  publisher = {American Physical Society},
  doi = {10.1103/PhysRevLett.125.073204},
  url = {https://link.aps.org/doi/10.1103/PhysRevLett.125.073204}
}

@article{Gonifmmode_CP_theory_PhysRevLett.131.186303,
  title = {Critical Phase Dualities in 1D Exactly Solvable Quasiperiodic Models},
  author = {Gon\ifmmode \mbox{\c{c}}\else \c{c}\fi{}alves, Miguel and Amorim, Bruno and Castro, Eduardo V. and Ribeiro, Pedro},
  journal = {Phys. Rev. Lett.},
  volume = {131},
  issue = {18},
  pages = {186303},
  numpages = {6},
  year = {2023},
  month = {Nov},
  publisher = {American Physical Society},
  doi = {10.1103/PhysRevLett.131.186303},
  url = {https://link.aps.org/doi/10.1103/PhysRevLett.131.186303}
}

@article{XIAO_Critical_MomentumLattice_20212175,
    title = {Observation of topological phase with critical localization in a quasi-periodic lattice},
    journal = {Science Bulletin},
    volume = {66},
    number = {21},
    pages = {2175-2180},
    year = {2021},
    issn = {2095-9273},
    doi = {https://doi.org/10.1016/j.scib.2021.07.025},
    url = {https://www.sciencedirect.com/science/article/pii/S2095927321005065},
    author = {Teng Xiao and Dizhou Xie and Zhaoli Dong and Tao Chen and Wei Yi and Bo Yan}
}

@article{Li_modifiedAAH_exp_2023,
    author={Li, Hao
    and Wang, Yong-Yi
    and Shi, Yun-Hao
    and Huang, Kaixuan
    and Song, Xiaohui
    and Liang, Gui-Han
    and Mei, Zheng-Yang
    and Zhou, Bozhen
    and Zhang, He
    and Zhang, Jia-Chi
    and Chen, Shu
    and Zhao, S. P.
    and Tian, Ye
    and Yang, Zhan-Ying
    and Xiang, Zhongcheng
    and Xu, Kai
    and Zheng, Dongning
    and Fan, Heng},
    title={Observation of critical phase transition in a generalized Aubry-Andr{\'e}-Harper model with superconducting circuits},
    journal={npj Quantum Information},
    year={2023},
    month={Apr},
    day={25},
    volume={9},
    number={1},
    pages={40},
    issn={2056-6387},
    doi={10.1038/s41534-023-00712-w},
    url={https://doi.org/10.1038/s41534-023-00712-w}
}

@article{Weitenberg_Simonet_2021_FlqReview, title={Tailoring quantum gases by Floquet engineering}, ISSN={1745-2473, 1745-2481}, url={https://www.nature.com/articles/s41567-021-01316-x}, DOI={10.1038/s41567-021-01316-x}, journal={Nature Physics}, author={Weitenberg, Christof and Simonet, Juliette}, year={2021}, month=aug, }

@article{Dotti_doubleLoc_PhysRevResearch.7.L022026,
  title = {Measuring a localization phase diagram controlled by the interplay of disorder and driving},
  author = {Dotti, Peter and Bai, Yifei and Shimasaki, Toshihiko and Dardia, Anna R. and Weld, David M.},
  journal = {Phys. Rev. Res.},
  volume = {7},
  issue = {2},
  pages = {L022026},
  numpages = {6},
  year = {2025},
  month = {Apr},
  publisher = {American Physical Society},
  doi = {10.1103/PhysRevResearch.7.L022026},
  url = {https://link.aps.org/doi/10.1103/PhysRevResearch.7.L022026}
}

@inbook{Aubry_1980, address={Dordrecht}, title={Metal-Insulator Transition in One-Dimensional Deformable Lattices}, ISBN={9789400990067}, url={http://link.springer.com/10.1007/978-94-009-9004-3_10}, DOI={10.1007/978-94-009-9004-3_10}, booktitle={Bifurcation Phenomena in Mathematical Physics and Related Topics}, publisher={Springer Netherlands}, author={Aubry, Serge}, editor={Bardos, Claude and Bessis, Daniel}, year={1980}, pages={163–184} }

@article{Wang_Steinberg_Jarillo-Herrero_Gedik_2013, title={Observation of Floquet-Bloch States on the Surface of a Topological Insulator}, volume={342}, ISSN={0036-8075, 1095-9203}, DOI={10.1126/science.1239834},  number={6157}, journal={Science}, author={Wang, Y. H. and Steinberg, H. and Jarillo-Herrero, P. and Gedik, N.}, year={2013}, month=oct, pages={453–457}}

@article{Uzan-Narovlansky_interbandBerryPhase, title={Observation of interband Berry phase in laser-driven crystals}, volume={626}, ISSN={0028-0836, 1476-4687}, DOI={10.1038/s41586-023-06828-5}, number={7997}, journal={Nature}, author={Uzan-Narovlansky, Ayelet J. and Faeyrman, Lior and Brown, Graham G. and Shames, Sergei and Narovlansky, Vladimir and Xiao, Jiewen and Arusi-Parpar, Talya and Kneller, Omer and Bruner, Barry D. and Smirnova, Olga and Silva, Rui E. F. and Yan, Binghai and Jim\'enez-Gal\'an, \'Alvaro and Ivanov, Misha and Dudovich, Nirit}, year={2024}, month=feb, pages={66–71} }

@article{Prosen_dimer_PhysRevLett.87.066601,
  title = {Dimer Decimation and Intricately Nested Localized-Ballistic Phases of a Kicked Harper Model},
  author = {Prosen, Toma\ifmmode \check{z}\else \v{z}\fi{} and Satija, Indubala I. and Shah, Nausheen},
  journal = {Phys. Rev. Lett.},
  volume = {87},
  issue = {6},
  pages = {066601},
  numpages = {4},
  year = {2001},
  month = {Jul},
  publisher = {American Physical Society},
  doi = {10.1103/PhysRevLett.87.066601},
  url = {https://link.aps.org/doi/10.1103/PhysRevLett.87.066601}
}

@article{Ketzmerick_Kruse_Geisel_1999_Lanzcos, title={Efficient diagonalization of kicked quantum systems}, volume={131}, rights={https://www.elsevier.com/tdm/userlicense/1.0/}, ISSN={01672789}, url={https://linkinghub.elsevier.com/retrieve/pii/S0167278998002309}, DOI={10.1016/S0167-2789(98)00230-9}, number={1–4}, journal={Physica D: Nonlinear Phenomena}, author={Ketzmerick, R. and Kruse, K. and Geisel, T.}, year={1999}, month=july, pages={247–253} }

@article{Borgonovi_Shepelyansky_1995_KHM, title={Spectral Variety in the Kicked Harper Model}, volume={29}, ISSN={0295-5075, 1286-4854}, url={https://iopscience.iop.org/article/10.1209/0295-5075/29/2/002}, DOI={10.1209/0295-5075/29/2/002}, number={2}, journal={Europhysics Letters (EPL)}, author={Borgonovi, F and Shepelyansky, D}, year={1995}, month=jan, pages={117–122} }

@article{Gonçalves_2024NatPhys, title={Incommensurability enabled quasi-fractal order in 1D narrow-band moiré systems}, volume={20}, ISSN={1745-2473, 1745-2481}, url={https://www.nature.com/articles/s41567-024-02662-2}, DOI={10.1038/s41567-024-02662-2}, number={12}, journal={Nature Physics}, author={Gonçalves, Miguel and Amorim, Bruno and Riche, Flavio and Castro, Eduardo V. and Ribeiro, Pedro}, year={2024}, month=dec, pages={1933–1940}}

@article{Arlinghaus_Holthaus_2010_OLstrongfieldSim, title={Driven optical lattices as strong-field simulators}, volume={81}, rights={http://link.aps.org/licenses/aps-default-license}, ISSN={1050-2947, 1094-1622}, url={https://link.aps.org/doi/10.1103/PhysRevA.81.063612}, DOI={10.1103/PhysRevA.81.063612}, number={6}, journal={Physical Review A}, author={Arlinghaus, Stephan and Holthaus, Martin}, year={2010}, month=june, pages={063612} }

\end{document}


\preprint{APS/123-QED}

\title{Supplementary Material: Exploring light-induced phases of 2D materials in a modulated 1D quasicrystal
}

\maketitle

\renewcommand\thefigure{S\arabic{figure}} 

\onecolumngrid
\tableofcontents

\newpage
\section{2D dynamics as a consequence of the cross-dimensional mapping}
\subsection{2D directional localization as a consequence of the cross-dimensional mapping}
The localization phase transition in the 1D model corresponds to freezing of 2D transport along the additional (physical) dimension when $\Delta < 2J$ ($\Delta > 2J$). The critical point $\Delta = 2J$ ($J_y = J$) corresponds to the isotropic case in 2D, where transport is allowed in both directions. This phenomenon in 2D is known as directional localization \cite{Barelli_PhysRevLett.83.5082, Itzler_Bojko_Chaikin_1992, Paul_PhysRevLett.132.246402}, and can be demonstrated through numerical integration as shown in Fig. 1 of the main text. Here we present additional numerical evidence and some comments regarding this phenomenon. 

\begin{figure}[htb]
    \includegraphics[width = \linewidth]{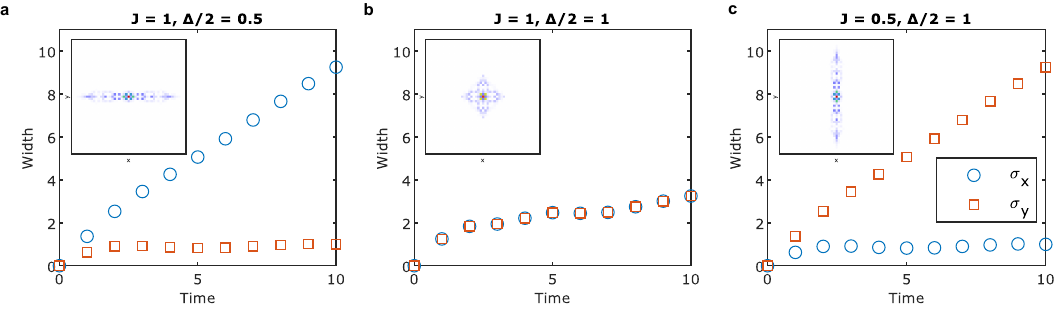}
    \caption{\textbf{2D directional localization for different strengths of anisotropy} $\Delta/J$. \textbf{a,} $\Delta/J = 1$ (ballistic phase in 1D). \textbf{b,} $\Delta/J = 2$ (critical point in 1D). \textbf{c,} $\Delta/J = 4$ (localized phase in 1D). Blue circles show the width in the $x$-direction and orange squares the width in the $y$-direction, defined as the second moment of the integrated density profile (Eq.\ref{eqS:2Dwidth}). Insets show the density profile of the time-evolved wave packet at the end of the simulation. }
    \label{figSI:dirLoc}
\end{figure}

Fig.~\ref{figSI:dirLoc} shows the results of numerical integration of the 2D Hofstadter Hamiltonian (Eq. \ref{eqS:harper}) by the Crank-Nicolson method ($\hbar = 1$) with $\beta = 1064/874.61$, a time step $\delta t = 0.01$ and $t_\mathrm{hold}$ up to 10. The system size is $50\times 50$, small enough to be aperiodic with respect to $\beta$, and the initial state is a single-site excitation at the grid center. Transport is measured by the width of the wave packet along the desired dimension $\sigma_{x,y}$, defined as the second moment of the density profile integrated over the orthogonal direction: 
\begin{equation}
    \sigma_{x} = \sqrt{\sum_{j=1}^L \left( (j-x_0)^2\sum_{l=1}^L\left\vert \Psi_{j,l}\right\vert^2 \right) }, \quad \sigma_{y} = \sqrt{\sum_{l=1}^L \left( (l-y_0)^2\sum_{j=1}^L\left\vert \Psi_{j,l}\right\vert^2 \right) }, \label{eqS:2Dwidth}
\end{equation}
where $x_0$ ($y_0$) is the center-of-mass position of the wave packet along that direction.

The numerical simulation cleanly demonstrates the phenomenon of directional localization. When the anisotropy favors the physical ($x$) dimension ($\Delta/J < 2$), the 1D system is in the delocalized ballistic phase, and the 2D wave packet expands ballistically along the $x$-direction but freezes along the $y$-direction after an initial transient expansion. When the system is isotropic ($\Delta/J = 2$), the 1D system is at the critical point and 2D expansion is allowed along both directions. When the ansisotropy favors the additional dimension ($\Delta/J>2$), the 1D system is in the localized phase, and the 2D wave packet stops expanding along the $x$-direction, but ballistically expands along the $y$-direction. The symmetry around the isotropic case is a simple consequence of the $90^\circ$ rotation symmetry in the 2D Hofstadter model, and represents a 2D consequence of the  Aubry duality in 1D \cite{aubry1980analyticity, Barelli_PhysRevLett.83.5082}. 

\subsection{Extending the mapping to driven systems}
We now briefly review the inclusion of driving by external irradiation in the 2D Hofstatder model, and its projection to the 1D model, which are discussed also in reference~\cite{Bai_PhysRevB.111.115163}. Assuming spatially homogeneous illumination, we choose the velocity gauge such that the electric field is completely specified by the vector potential: 
\begin{align*}
    &\mathbf{A} = \mathbf{A}_B + \mathbf{A}_E, \quad \mathbf{A}_B = (0,2\pi\beta j, 0), \\
    & \mathbf{A}_E = \left(K_0 \sin (\omega t + \theta_x), \, \varphi_0 \sin (\omega t + \theta_y), \, 0 \right). 
\end{align*}
Here $K_0$ and $\varphi_0$ describe the strength of the electric field along the physical and additional dimension, respectively. The polarization of the irradiation is captured in the phase difference $\vartheta = \theta_y - \theta_x$.

Through the Peierls substitution, the 2D driven Harper-Hofstadter model can be written
\begin{equation*}
    \hat H_\mathrm{DHH} = \sum_{j,l} -J e^{-iK_0\sin(\omega t + \theta_x)}\hat d_{j+1, l}^\dagger \hat d_{j,l} + \frac{\Delta}{2} e^{-i (2\pi\beta j + \varphi_0 \sin(\omega t + \theta_y)} \hat d_{j, l+1}^\dagger \hat d_{j,l} + \mathrm{h.c.}.
\end{equation*}
The same method used in the static case still applies here, since the Hamiltonian remains independent of the additional dimension \cite{Bai_PhysRevB.111.115163}. We thus arrive at the 1D Hamiltonian studied in this work, reproduced here for convenience:
\begin{align}
    \hat H (t) = &\sum_j -J e^{-iK_0 \sin (\omega t + \theta_x)}\hat c_{j+1}^\dagger \hat c_j + \mathrm{h.c.} + \Delta \cos(2\pi\beta j + \kappa + \varphi_0\sin(\omega t + \theta_y) ) \hat n_j . \label{eqS:DDAAH_comoving}
\end{align}
Crucially, the electric field component along the additional dimension in 2D maps to the phasonic modulation in the 1D system~\cite{Shimasaki_PhysRevLett.133.083405, Bai_PhysRevB.111.115163}. We refer to this Hamiltonian as the doubly-driven AAH (DDAAH) model.

\section{Spectral and localization properties of the doubly-driven AAH model}
In this section, we review and provide additional details on the spectral and localization properties of the DDAAH model, relevant to the experimental setting of the main text. More discussions can be found in \cite{Bai_PhysRevB.111.115163}.

\subsection{Floquet analysis}
The theoretical and numerical analyses shown in this section rely on the Floquet formalism for a time-periodic Hamiltonian, of which we provide a short overview here. By the Floquet theorem, there exists a set of states forming an orthonormal and complete basis for a time-periodic Hamiltonian $\hat H(t) = \hat H(t+T)$ (where $T=2\pi/\omega$ is the driving period). States $\vert \psi_n (t)\rangle$ in this basis (where $n$ labels the state) are referred to as the Floquet eigenstates, and time-evolve as $\vert \psi_n (t+T)\rangle = e^{-i \varepsilon_n T/\hbar} \vert \psi_n (t)\rangle$ where $\varepsilon_n$ is the quasienergy. In an exact analogy to Bloch's theorem for spatially periodic systems, the Floquet eigenstate can be decomposed into a plane-wave factor and a time-periodic Floquet function $\vert \Psi_n (t)\rangle$: 
\[
\vert \psi_n (t)\rangle = e^{-i\varepsilon_n t/\hbar} \vert \Psi_n (t)\rangle , \quad  \vert \Psi_n (t+T)\rangle = \vert \Psi_n (t)\rangle. 
\]
Because the Floquet function $\vert \Psi_n (t)\rangle$ is time-periodic, it can be expanded by a discrete Fourier series as $\vert \Psi_n (t)\rangle = \sum_{m=-\infty}^\infty \vert \phi_{n,m} \rangle e^{im\omega t}$ where $m$ labels the Fourier harmonics. In combination with the discrete Fourier expansion of the Hamiltonian $\hat H(t) = \sum_{m=-\infty}^\infty \hat H_m e^{im\omega t}$, an eigenvalue equation is derived for the $m$-th harmonics of the Floquet function
\[
\left(\varepsilon_n + m\hbar \omega \right) \vert \phi_{n,m} \rangle = \sum_{m'}\hat H_{m-m'} \vert \phi_{n,m'} \rangle.
\]
This equation can be rewritten as 
\begin{equation}
    \mathcal{H}\varphi_n = \varepsilon_n \varphi_n, \quad \mathcal{H} = \begin{pmatrix}
    \ddots & \hat H_{-1} & \hat H_{-2} & \hat H_{-3} \\
    \hat H_{1} & \hat H_0-m\hbar \omega & \hat H_{-1} & \hat H_{-2} \\
    \hat H_{2} & \hat H_{1} & \hat H_0 - (m+1)\hbar \omega & \hat H_{-1} \\
    \hat H_3 & \hat H_2 & \hat H_1 & \ddots
    \end{pmatrix}, \quad \varphi_n = \begin{pmatrix}
    \vdots \\
    \vert \phi_{n,m} \rangle \\
    \vert \phi_{n,m+1} \rangle \\
    \vdots
    \end{pmatrix} \label{eq:flq_blockH}
\end{equation}
We can now straightforwardly obtain properties of the time-periodic Hamiltonian $\hat H(t)$ through this eigenvalue equation, such as its quasienergy spectrum, and the spatial distribution of the Floquet eigenstates. 

The Fourier modes of our model of interest, the DDAAH Hamiltonian, can be written as
\begin{align*}
    \hat H_m = \sum_{j=1}^{L-1} -e^{im \theta_x} J \mathcal{J}_m (K_0)\left((-1)^m\hat c_{j+1}^\dagger \hat c_j + \hat c_{j}^\dagger \hat c_{j+1}\right) + \sum_{j=1}^{L} e^{im \theta_y}\Delta  \mathcal{J}_m (\varphi_0)i^m \cos \left(2\pi\beta j +\kappa - \frac{m\pi}{2} \right)\hat n_j, 
\end{align*}
which can be straightforwardly derived from the Jacobi-Anger expansion \cite{Bai_PhysRevB.111.115163}. 

\subsection{High frequency expansion and the effective Hamiltonian}
The high-frequency expansion \cite{Goldman_PhysRevX.4.031027, Eckardt_2015_HFE} is a useful method to derive the effective Hamiltonian $\hat H_\mathrm{eff}$ associated with a Floquet Hamiltonian $\hat H(t)$ in the high-frequency regime. The expansion gives 
\begin{equation*}
    \hat H_\mathrm{eff} =\sum_{n=0}^{\infty} \hat H^{(n)} =\hat H^{(0)} + \hat H^{(1)} + \hat H^{(2)} + \mathcal{O}\left((\hbar\omega)^{-3} \right). \label{eq:HFE}
\end{equation*}
Here we primarily focus on the first two terms
\begin{align*}
    \hat H^{(0)} = \frac{1}{T}\int^{T/2}_{-T/2}\mathrm{d}t \, \hat H(t) = \hat H_0, \quad \hat H^{(1)} = \sum_{m=1}^\infty \frac{\left[ \hat H_m ,\hat H_{-m}\right]}{m\hbar \omega}.
\end{align*}
We observe that $\hat H^{(0)}$ is simply the time-averaged term of $\hat H(t)$ over a period, whereas the term $\hat H^{(1)}$ describes the virtual $m$-photon absorption and re-emission processes \cite{Kitagawa_PhysRevB.84.235108}. The expression for $\hat H^{(2)}$ involves the effect of higher-harmonics and harmonic mixing. 

From the expression of $\hat H_m$, we have
\begin{align}
    & \hat H^{(0)} = \sum_j -J \mathcal{J}_0 (K_0) \left(\hat c_{j+1}^\dagger \hat c_j + \mathrm{h.c.}\right) + \Delta \mathcal{J}_0 (\varphi_0) \cos \left(2\pi\beta j + \kappa \right)\hat n_j \\
    & \hat H^{(1)} = \frac{4J\Delta}{\hbar \omega} \sin(\pi\beta) \sum_{m=1}^\infty \frac{i^{m+1}}{m}\sin\left(m \vartheta \right)\mathcal{J}_m(K_0)\mathcal{J}_m (\varphi_0)\sum_{j=1}^{L-1}\sin\left(2\pi\beta j + \kappa + \pi\beta -\frac{m\pi}{2}\right) \left( -\hat c_{j+1}^\dagger \hat c_j +(-1)^m \hat c_{j}^\dagger \hat c_{j+1} \right). \label{eqSI:h1}
\end{align}
In other words, $\hat H^{(0)}$ is the regular Aubry-Andr\'e-Harper model with the tunneling and quasiperiodic strengths rescaled by Bessel functions of the corresponding field amplitudes. $\hat H^{(1)}$ on the other hand represents a quasiperiodically modulated nearest-neighbor tunneling, which supports a critical phase hosting multifractal wave functions \cite{Han_cri_bicrit_modifiedAAH_PhysRevB.50.11365, Chang_MFanalysis_modifiedAAH_PhysRevB.55.12971}. The interference term $\sin\left(m \vartheta \right) = \sin\left(m (\theta_y - \theta_x) \right)$ in $\hat H^{(1)}$ provides a selection rule for the virtual photon processes: $\hat H^{(1)} = 0$ when the polarization is linear ($\vartheta = 0$), and only the odd $m$ terms contribute when $\vartheta = \pm \pi/2$. We further evaluated the next order term $\hat H^{(2)}$, though the result is too lengthy to be displayed here. In short, it \rr{is non-vanishing for either polarization,} contains quasiperiodically modulated next-nearest-neighbor tunneling in 1D, and additional quasiperiodic potential. 

We now briefly discuss the effective Hamiltonian from the perspective of the higher-dimensional space. In the example of a Floquet topological insulator formed by irradiated graphene, it is precisely this term $\hat H^{(1)}$ that generates the next nearest-neighbor tunneling term in the real space and leads to an effective Haldane model~\cite{Kitagawa_PhysRevB.84.235108}. In our case, from the perspective of the higher-dimensional space in which the Harper model $\hat H_\mathrm{Harper}$ lives, the term $\hat H^{(1)}$ also generates an effective next nearest-neighbor tunneling when the polarization is not linear. When this effective model is projected to the 1D space, this effective next nearest-neighbor tunneling maps to a quasiperiodically modulated nearest-neighbor tunneling \cite{Bai_PhysRevB.111.115163}.  

\subsection{Dependence of anomalous expansion on quasiperiodic strength and driving frequency}
\begin{figure}[h!]
    \centering
    \includegraphics[width = 180mm]{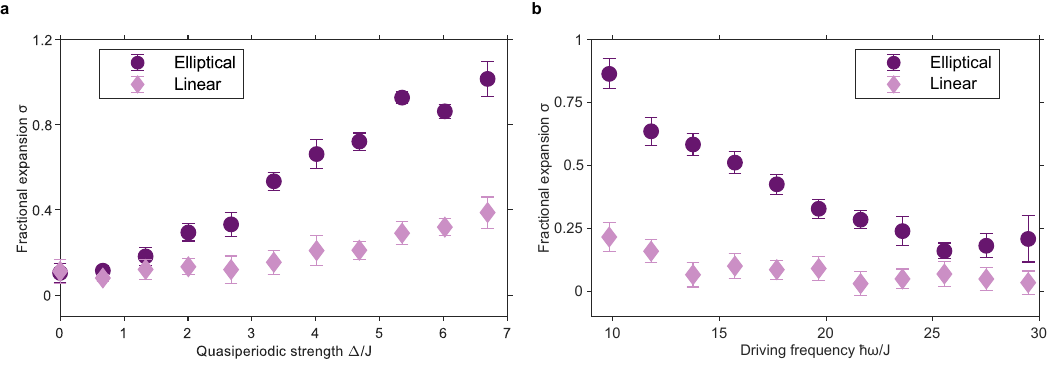}
    \caption{\rr{
    \textbf{Dependence of anomalous expansion on quasiperiodic strength and driving frequency at $K_0 = \varphi_0 = 2.405$.} \textbf{a} Expansion versus quasiperiodic strength $\Delta$ at $\hbar\omega \approx 9.83J$. \textbf{b} Expansion versus driving frequency $\hbar\omega$ at $\Delta \approx 5.01J$. Small deviations from expected behaviors can be attributed to experimental imperfections and increasingly significant contributions from $\hat H^{(2)}$ at larger $\Delta$.
    }
    }
    \label{figSI:s2andhwdep}
\end{figure}
\rr{
The strength of $\hat H^{(1)}$ that leads to the multifractal phase increases (decreases) as $\Delta$ ($\hbar\omega$) increases respectively, due to the prefactor $\Delta/\hbar\omega$ in Eq.(\ref{eqSI:h1}). Motivated by this relation, in Figure \ref{figSI:s2andhwdep} we show experimental measurements of fractional expansion varying the quasiperiodic strength $\Delta$ and driving frequency $\hbar\omega$, taken at $K_0=\varphi_0 = 2.405$ where the time-averaged Hamiltonian $\hat H^{(0)}$ vanishes completely. 
}

\rr{
The observed dependence of anomalous expansion follows the expected behaviors for both cases. For linearly polarized driving, we observe barely any expansion in either case other than at very large quasiperiodic strength $\Delta/J$, at which point $\hat H^{(2)}$ starts to contribute. For elliptically polarized driving, we observe that the expansion increases as $\Delta$ increases and falls as $\hbar\omega$ increases, matching closely to the expected behaviors because the only effective kinetic energy comes from $\hat H^{(1)}$. These observations further support our claims of a nontrivial multifractal phase originating from $\hat H^{(1)}$. 
}

\subsection{Quasienergy, Floquet eigenstates and multifractality}
\subsubsection{Methods and diagnostics}
In this section, we will discuss the quasienergy spectra and the multifractality characterisics of the Floquet eigenstates. We focus on the specific case studied in the main text, in which $K_0 \approx 2.405$ such that there is effectively no tunneling at the zeroth order. We keep $\Delta/J \approx 5.01$ and $\hbar\omega = h\times 500\, \mathrm{Hz} \approx 9.82J$, in accordance with the experiment. The quasienergy spectrum is obtained by diagonalizing the block Hamiltonian (Eq. (\ref{eq:flq_blockH})), keeping only the central Floquet-Brillouin zone. The multifractality characteristics are calculated from the generalized inverse participation ratio (IPR) in the lattice basis. For a normalized wave function $\psi_j$, it is defined as
\begin{equation}
    \mathrm{IPR}_q = \sum_{j=1}^L \vert \psi_j \vert^{2q}. \label{eq:IPR}
\end{equation}
The standard IPR corresponds to $q = 2$. For a localized state, $\mathrm{IPR}_q$ is nonvanishing; for a delocalized (extended or critical) state, $\mathrm{IPR}_q \rightarrow 1/L^{q/2}$. Here we focus on $q>0$. The multifractal nature of a wave function is further probed by calculating the multifractal dimension $D_q$
\begin{equation}
    \mathrm{IPR}_q \sim L^{-\tau_q}, \quad D_q = \frac{\tau_q}{q-1}.
\end{equation}
For a localized state, $D_q = 0$; an extended state has $D_q = 1$ and a multifractal state has an intermediate value of $D_q$ for all $q>0$. 

\subsubsection{Quasienergy}
\begin{figure}[h!]
    \centering
    \includegraphics[width = 180mm]{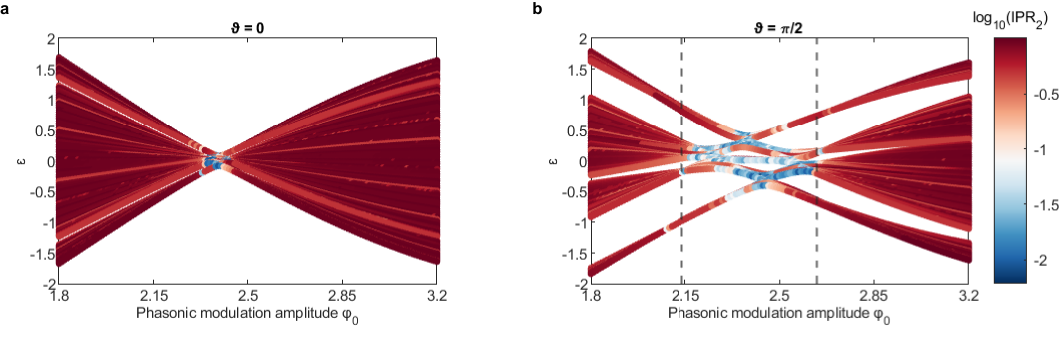}
    \caption{\textbf{Comparison of quasienergy spectra under elliptically and linearly polarized modulation}. 
    Quasienergy spectra as a function of phasonic modulation amplitude for linearly (left) and elliptically (right) polarized modulation. For both panels $K_0 = 2.405$, $\Delta/J \approx 5.01$ and $\hbar\omega = h\times 500\, \mathrm{Hz} \approx 9.82J$. Dashed lines mark the predicted phase transition from localized to critical phases based on the effective Hamiltonian $\hat H^{(0)} + \hat H^{(1)}$ only. Colormap represents the logarithm of the $\mathrm{IPR}_2$: localized Floquet eigenstates have $\log_{10}(\mathrm{IPR}_2) \approx 0$, extended eigenstates have $\log_{10}(\mathrm{IPR}_2) \approx -2$, and critical eigenstates lie between these extremes. Both calculations are done with a rational approximant $\beta = 309/254$ and $L = 254$ with periodic boundary conditions. }
    \label{fig:QE_DL_VaryPhi}
\end{figure}

The quasienergy spectra for linearly and elliptically polarized modulations, shown in Fig.\ref{fig:QE_DL_VaryPhi}, are drastically different. For the case of linear polarization, the spectrum shows very small gaps, since the effective quasiperiodic strength diverges as $K_0$ is near the zeros of the Bessel function, and all eigenstates are localized for all plotted phasonic amplitudes $\varphi_0$. The bandwidth also vanishes almost completely when $\varphi_0 \approx 2.405$ where $\hat H_\mathrm{eff} \approx \hat H^{(0)} \approx 0$. 

In contrast, the spectrum under elliptically polarized modulations shows much larger gaps and the bandwidth does not vanish near $\varphi_0 \approx 2.405$ due to $\hat H^{(1)}$. The eigenstates near $\varphi_0\approx 2.4$ show non-trivial localization properties, captured by $\log_{10}(\mathrm{IPR}_2)$. The numerical values of $\log_{10}(\mathrm{IPR}_2)$ suggest that the states are neither localized nor fully extended. We draw the boundary between multifractal and localized phases predicted by considering only the first two terms in the high-frequency expansion $\hat H^{(0)} + \hat H^{(1)}$, and find that they show excellent agreement with the observed change in the multifractal properties as we vary the phasonic amplitude $\varphi_0$.

\subsubsection{Multifractality of Floquet eigenstates}
\begin{figure}[h!]
    \centering
    \includegraphics[width = 180mm]{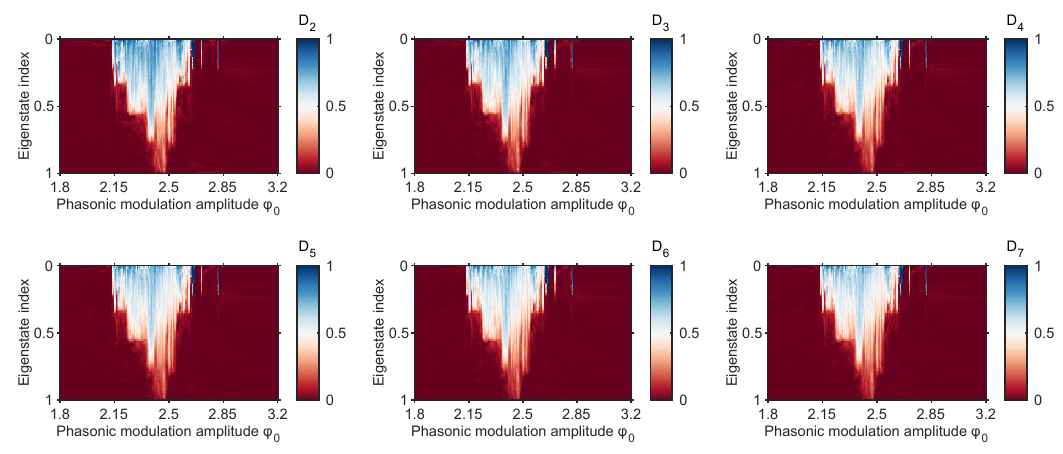}
    \caption{\textbf{Generalized fractal dimension $D_q$ of Floquet eigenstates as a function of the phasonic modulation amplitude $\varphi_0$ at $K_0 = 2.405$.}
    $D_q$ shows nontrivial structure over a range of different values of $q$, from $q =2$ to $q = 7$. Here $K_0 = 2.405$, $\Delta/J \approx 5.01$ and $\hbar\omega = h\times 500\, \mathrm{Hz} \approx 9.82J$. The system size is varied from $L = 100$ sites to $900$ sites. }
    \label{fig:Dq}
\end{figure}

While the fractal dimension $D_2$ is a relevant metric characterizing eigenstates, rigorously speaking one needs to calculate $D_q$ for different $q$ to establish multifractality. Here we provide the results of such a calculation. We focus on the first anomalous region ($\varphi_0 \in [1.8,3.2]$) with $K_0 = K_0^\mathrm{DL} \approx  2.405$, and calculate $D_q$ up to $q = 7$. The numerical results shown in Fig.\ref{fig:Dq} cleanly demonstrate the multifractality of the Floquet eigenstates. 

\section{Interactions and heating}
At the temperatures used in this work, the ${}^{84}\mathrm{Sr} - {}^{84}\mathrm{Sr}$ interaction is characterized by the $s$-wave scattering length of $123a_0$ where $a_0$ is the Bohr radius. Since the radial confinement is weak (see Method), our system is in the weakly-interacting, mean-field regime. Here we discuss the consequences of these weak interactions in our experiments. 

The weak radial confinement implies that the dominant heating mechanism is transverse excitation due to collision \cite{Chaudhury_transverseInstability_PhysRevA.91.023624}. The low overall trap depth sets a limit on the maximal possible temperature attainable through such thermalization. Indeed, we have experimentally observed through time-of-flight images that the maximal transverse energy of the condensate saturates early in the evolution, instead of increasing indefinitely under the periodic modulation, and remains saturated until the longest evolution time of $t_\mathrm{hold} = 20\, \mathrm{s}$. Rather than an increase in temperature, the heating then manifests itself as one-body losses, which act to reduce the temperature evaporatively \cite{Reitter_heating_PhysRevLett.119.200402}. We did not observe any population in the higher band during the entirety of the time evolution, suggesting that heating rates due to transitions to higher bands are minimal \cite{Weinberg_multiphotonTransition_PhysRevA.92.043621}, and justifying the use of a single-band tight-binding approach. 

The Bose-Hubbard interaction parameter $U$ can be estimated as $U = (4\pi\hbar^2 a_s/m)\int \vert \psi(\mathbf{r})\vert^4 \mathrm{d}^3 \mathbf{r}$ where $m$ is the mass of the ${}^{84}\mathrm{Sr}$ atom and $\psi(\mathbf{r})$ is the single-site wave function. We take a Gaussian approximation for $\psi(\mathbf{r})$, using the measured \textit{in situ} transverse width of about $6.5\, \mathrm{\mu m}$. This gives $U \approx 0.0015J$ which provides an upper bound on the interaction energy scale. Atom loss during an experiment further reduces the interaction, and thus slows the thermalization rate. 

During a previous related study \cite{shimasaki2022anomalous}, we observed experimentally and numerically that the basic structure of the phase diagram is unchanged over a large range of interaction strengths. Details of this complex, non-equilibrium interplay between localization, external driving, interaction and losses are extremely challenging to simulate numerically given our long evolution times, and are beyond the scope of this work. Rather, we experimentally benchmark our expansion dynamics up to the longest possible evolution time of $t_\mathrm{hold} = 20\,\mathrm{s} \approx 6400 T_J$. \rr{The results are shown in  Appendix D in the main text.} 

\section{Dependence of initial width on quasiperiodic strength}
\begin{figure}[h!]
    \centering
    \includegraphics[width = 89mm]{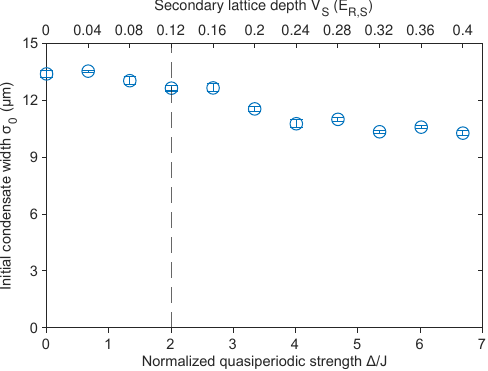}
    \caption{\textbf{The widths along the lattice direction of the condensate $\sigma_0$ at the beginning of the experiment}. The dashed line marks the expected critical quasiperiodicity strength $\Delta/J =2$. Error bars represent standard error of the mean of 5 repeated measurements. }
    \label{figSI:initialWidthVSs2}
\end{figure}

Our loading procedure means that the initial width depends weakly on the depth of the secondary lattice $V_S$, ergo on the quasiperiodic potential strength $\Delta$. During the loading, the condensate expands slightly along the lattice direction as the crossed dipole trap weakens. This expansion is slower for increasing quasiperiodic strength. To calibrate this effect, we measured the initial width at different quasiperiodic strengths $\Delta$ under the same experimental conditions as those in the main text. The result is shown in Fig. \ref{figSI:initialWidthVSs2}. In practice, we measured the initial condensate width after each data sets taken under the same experimental condition, which were used to compute the fractional width $\sigma = \sigma_x/\sigma_0$ used in the main text. 

\subsection*{Calibration of the dimensionless modulation amplitudes for each polarization}
We calibrated the dimensionless modulation amplitudes $K_0$ and $\varphi_0$ based on the expansion of the condensate using dynamic localization (DL) \cite{Lignier_DL_PhysRevLett.99.220403} and reversible phasonic control of localization (RPCL) \cite{Shimasaki_PhysRevLett.133.083405}. As discussed in the main text, DL refers to the coherent renormalization of the tunneling energy $J$ by the dipolar amplitude $K_0$, and RPCL refers to the coherent renormalization of the quasiperiodic disorder $\Delta$ by the phasonic amplitude $\varphi_0$. Thus, we use DL to calibrate $K_0$ and RPCL to calibrate $\varphi_0$.
We remark that the close resemblance between the theoretical and experimental phase diagrams without any fitting parameters (Fig. 2 in the main text) illustrates the accuracy of mapping dimensionless modulation amplitudes from frequency modulations, for the case of elliptically polarized driving. 

\begin{figure}[h!]
    \centering
    \includegraphics[width = 89mm]{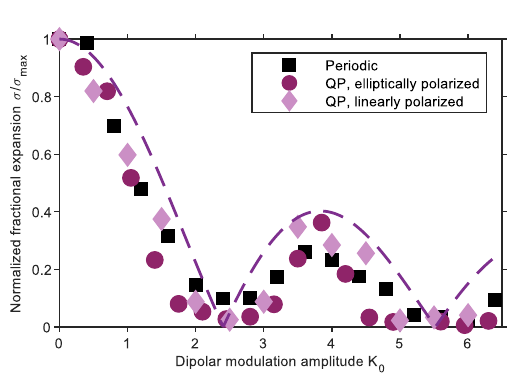}
    \caption{\textbf{Calibration of dipolar modulation amplitude} $K_0$ for the purely periodic case ($V_S = 0$, black square marker), and for elliptically (dark diamond markers) and linearly (light circle marker) polarized modulations. The dashed curve is $\vert\mathcal{J}_0(K_0)\vert$ without any fit parameter. For the periodic case, $t_\mathrm{hold} = 1000\,\mathrm{ms}$. QP: quasiperiodic cases with $V_S = 0.3E_{R,S}$. For the elliptically polarized case, $\varphi_0 = 2.45$ (near the first zero of the Bessel function) and $t_\mathrm{hold} = 1250\,\mathrm{ms}$; for the linearly polarized case, $\varphi_0 = 2.405$ (near the first zero of the Bessel function) and $t_\mathrm{hold} = 1000\mathrm{ms}$. For all data here, $\omega = 2\pi\times 500\,\mathrm{Hz}$ and $V_P = 9E_{R,P}$. }
    \label{fig:K0_calibration}
\end{figure}

The expressions for the dipolar amplitude $K_0$, and thus $f_{0,P}$, are independent of the polarizations and $\varphi_0$. We experimentally confirmed this point (Fig. \ref{fig:K0_calibration}). All data follows the shape of the absolute value of the Bessel function $\vert \mathcal{J}_0(K_0) \vert$.  

\begin{figure}[h!]
    \centering
    \includegraphics[width = 180mm]{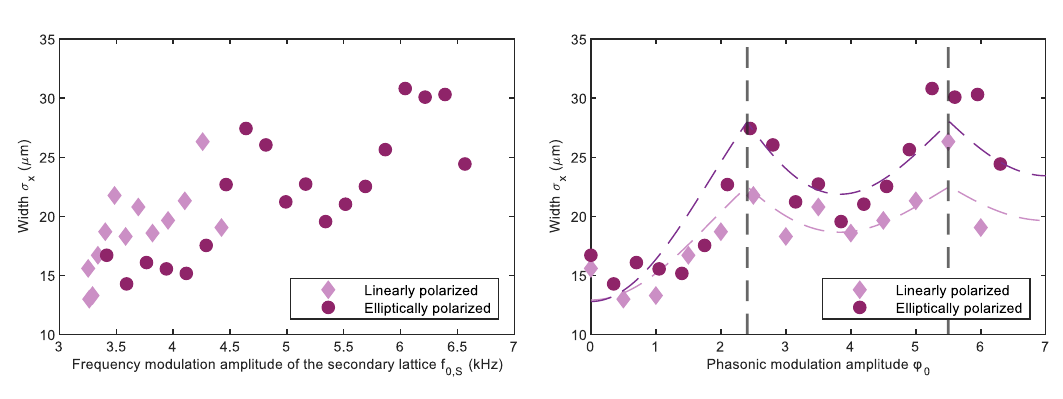}
    \caption{\textbf{Calibration of the phasonic modulation amplitude} using reversible phasonic control of localization \cite{Shimasaki_PhysRevLett.133.083405}  
    for linearly (circles) and elliptically (diamonds) polarized modulation. \textbf{Left,} condensate width versus frequency modulation amplitude of the secondary lattice $f_{0,S}$. \textbf{Right,} condensate width versus phasonic modulation amplitude, after converting $f_{0,S}$ to the dimensionless driving parameter $\varphi_0$, according to Eqs.(\ref{eq:phasonAmpEP}) and (\ref{eq:phasonAmpLP}). Vertical lines mark the first two zeros of the Bessel function $\mathcal{J}_0 (\varphi_0)$. For the linearly polarized case, $K_0 = 1$, $V_S = 0.3E_{R,S}$ and $t_\mathrm{hold} = 500\,\mathrm{ms}$. For the elliptically polarized case, $K_0 =1.05$, $V_S = 0.28 E_{R,S}$ and $t_\mathrm{hold} = 1250\,\mathrm{ms}$. For all data shown here, $\omega = 2\pi\times 500\,\mathrm{Hz}$ and $V_P = 9E_{R,P}$. Dashed curves are inverted Bessel functions, scaled only vertically, serving as guides to the eye.}
    \label{fig:varphi_EPLP_Calibration}
\end{figure}

More care needs to be taken for $\varphi_0$, because the values of frequency modulation amplitude $f_{0,S}$ are different for linear or elliptical polarization for the same $\varphi_0$. This is explicitly shown in Fig.\ref{fig:varphi_EPLP_Calibration}(a). However, these two cases follow the same trend as observed in \cite{Shimasaki_PhysRevLett.133.083405} once the dimensionless phasonic amplitude $\varphi_0$ is extracted accordingly, shown in Fig.~\ref{fig:varphi_EPLP_Calibration}(b). The peaks also matches the zeros of the Bessel function $\mathcal{J}_0(\varphi_0)$ where maximal expansion is expected. 

From these data we set and calibrate the values of the modulation parameters discussed in the main text.

\bibliography{suppInfoBib}